\documentstyle[emulateapj,apjfonts,epsfig]{article}

\newcommand{\ta}{triple-$\alpha$ }
\newcommand{\fc}{$F_{\rm crust}$ }

\begin{document}

\normalsize

\title{Unstable Nonradial Oscillations on Helium Burning Neutron Stars}

\author{Anthony L. Piro}
\affil{Department of Physics, Broida Hall, University of California
	\\ Santa Barbara, CA 93106; piro@physics.ucsb.edu}

\and

\author{Lars Bildsten}
\affil{Kavli Institute for Theoretical Physics and Department of Physics,
Kohn Hall, University of California
	\\ Santa Barbara, CA 93106; bildsten@kitp.ucsb.edu}

\begin{abstract}

  Material accreted onto a neutron star can stably burn
in steady state only when the accretion rate is high
(typically super-Eddington) or if a large flux from the
neutron star crust permeates the outer atmosphere. For
such situations we have analyzed the stability of nonradial
oscillations, finding one unstable mode for pure helium
accretion. This is a shallow surface wave which resides
in the helium atmosphere above the heavier ashes of the ocean. It is
excited by the increase in the nuclear reaction rate during the
oscillations, and it grows on the timescale of a second. For a
slowly rotating star, this mode has a frequency
$\omega/(2\pi)\approx (20-30\textrm{ Hz})(l(l+1)/2)^{1/2}$, and we
calculate the full spectrum that a rapidly rotating
($\gg30\textrm{ Hz}$) neutron star would support. The short
period X-ray binary 4U 1820--30 is accreting helium rich material
and is the system most likely to show this unstable mode, especially
when it is not exhibiting X-ray bursts. Our discovery of an
unstable mode in a thermally stable atmosphere shows that nonradial
perturbations have a different stability criterion than the spherically
symmetric thermal perturbations that generate type I X-ray bursts.

\end{abstract}


\section{Introduction}

  Type I X-ray bursts are caused by unstable nuclear burning on
accreting neutron stars (NSs) (see recent reviews by Bildsten 1998;
Strohmayer \& Bildsten 2003) in low mass X-ray binaries (LMXBs).
They have rise times of seconds with decay times ranging from tens
to hundreds of seconds depending on the thickness and composition
of the accumulated material. These bursts repeat every few hours
-- the timescale to accumulate an unstable amount of fuel on a NS.
Evidence that nuclear burning powers the bursts is the observed
ratio of the time averaged accretion luminosity to the time averaged
burst luminosity ($\approx 40$ for hydrogen rich accretion and
$\approx 125$ for pure helium accretion). The thermonuclear
instability which triggers these bursts is usually due to the
extreme temperature sensitivity of \ta reactions (Hansen \& Van Horn
1975; Woosley \& Taam 1976; Maraschi \& Cavaliere 1977; Joss 1977;
Lamb \& Lamb 1978).

  Ever since the original work of Hansen \& Van Horn (1975), the
approach to assessing the stability of burning on a NS has been to
apply a spherically symmetric thermal perturbation to a spherical model
(Schwarzschild \& H\"{a}rm 1965). This is either done numerically
through a time-dependent code (e.g. Woosley et al. 2003),
analytically with a one-zone model (e.g. Fujimoto, Hanawa \& Miyaji 1981),
or using an eigenvalue approach (e.g. Narayan \& Heyl 2003).
These results roughly agree and find a thermal runaway once a
critical amount of fuel has accumulated. These calculations also find
the accretion rate above which burning is stable.

  We are now taking an alternative approach to this problem. Using a
NS envelope that is stable to spherically symmetric thermal perturbations
(i.e. not bursting), we ask whether the envelope is unstable to a
non-axisymmetric perturbation. Since the immediate response to
such a perturbation is a wave, the proper way to proceed is through a
study of the nonradial oscillation modes. The criterion for thermal
versus mode stability is mathematically different (as we will explain
in \S2.2 and \S3.3), but it is not yet clear how this contrast will
manifest itself observationally. Nonradial oscillations in burning NS
envelopes have been studied before (for example McDermott \& Taam 1987;
Strohmayer \& Lee 1996), but these are for atmospheres that are thermally
unstable, leaving our question unanswered.

  To simplify this initial work, we assume that the NS is accreting
pure helium and burning this fuel at the same rate it accretes. This
steady state burning proceeds if the NS envelope is thermally stable
(and no type I X-ray bursts are occuring). It is stable when either the
temperature sensitivity of \ta reactions has weakened due to a high
envelope temperature ($\gtrsim5\times10^8$ K), or the thermal balance
of the burning layer is fixed by a high flux from deeper parts of the NS
(Paczy\'{n}ski 1983a). We then find the eigenfrequencies and test the
stability of g-modes in such an envelope.
The one unstable mode we find is a shallow surface
wave in the helium layer that is supported by the buoyancy of the
helium/carbon interface. It has a non-rotating frequency of
$\omega/(2\pi)\approx (20-30\textrm{ Hz})(l(l+1)/2)^{1/2}$,
depending on the accretion rate.
This mode
is driven unstable by the $\epsilon$-mechanism (i.e. pumped by the
perturbation in nuclear burning) and grows in amplitude on the timescale
of a second. This is very different from other observed stellar pulsations,
such as Cepheid variables, which are driven by the $\kappa$-mechanism
(pumped by changes in opacity). Unlike the damped higher radial order
g-modes, the unstable mode suffers little radiative damping in the outer
atmosphere. Rapid rotation like that seen from many accreting NSs
(200-700 Hz) will yield multiple oscillation frequencies with a well
defined pattern. If such a frequency pattern were seen from a NS it
would give strong evidence that a mode has been excited.

  In \S 2 we describe the physics of a NS star envelope accreting
and burning helium in steady state. The conditions required for steady
state accretion are reviewed. Nonradial adiabatic oscillations in
a plane-parallel envelope are discussed in \S 3, and though much
of this has been described in previous papers, we include
additional details and strategies that are relevant for NSs. The
oscillatory equations are derived in \S 3.1, and in \S 3.2 we
calculate the radial eigenfunctions and find the g-mode frequencies
for a non-rotating NS. The stability of these modes is investigated
in \S 3.3, including a look at excitation and damping times.
In \S 3.4 we find the unstable mode's frequency on a rapidly rotating
neutron star. The LMXB 4U 1820--30 is discussed in \S 4 along with other
candidates that might show the excited mode we have found. We conclude
in \S 5 by summarizing our findings and discussing future directions
for this work.

\section{Steady State Helium Accretion}

  In solving for a NS envelope we closely follow the model described
by Brown \& Bildsten (1998). We assume that pure helium is accreting
onto a NS and burning into heavier elements. It initially burns into
$^{12}$C via the \ta process and then heavier elements ($^{16}$O,
$^{20}$Ne, $^{24}$Mg, and $^{28}$Si) via subsequent alpha captures.
The envelope consists of three layers; a layer of pure accreted helium
which is the ``atmosphere,'' a thin layer where the burning occurs,
and a layer of heavy element ashes, the ``ocean.'' Below this is the
rigid NS crust which will provide the bottom boundary for the modes.

\subsection{Finding the Envelope Profile}

  We assume that the NS has a mass of $M = 1.4 M_\odot$ and a
radius of $R=10\textrm{ km}$. Neglecting general relativity, the
surface gravitational acceleration is then
$ g = GM/R^2 \approx 1.87\times 10^{14} \textrm{ cm s}^{-2}$.
The pressure scale height is $h = P/\rho g \approx 200 \textrm{ cm}$
at the burning layer. Since $h\ll R$ we assume that $g$ is constant
and that the envelope has plane-parallel geometry. This simplifies
calculations of the envelope profile and the nonradial perturbations
considered in \S 3. Hydrostatic balance then yields $P=gy$, where $y$
is the column depth, $dy\equiv-\rho dr$, and $r$ is the
radial distance.

  The plane-parallel nature of this problem implies that stability should
depend on the accretion rate per unit area, $\dot{m}$
(Fujimoto, Hanawa \& Miyaji 1981; Fushiki \& Lamb 1987b).
We therefore parametrize the accretion in terms of the local
Eddington rate per unit area for helium accretion (henceforth, simply
referred to as the Eddington rate),
\begin{eqnarray}
        \dot{m}_{\rm Edd} = \frac{2m_pc}{R\sigma_{\rm Th}}
        = 1.5\times10^5\textrm{ g cm}^{-2}\textrm{ s}^{-1}
                        \left(\frac{10^6\textrm{ cm}}{R}\right),
\end{eqnarray} 
where $m_p$ is the proton mass, $c$ is the speed of light,
and $\sigma_{\rm Th}$ is the Thomson cross section. NSs can accrete locally
at rates higher than this if the accretion geometry allows radiation
to escape without impeding the accretion flow. The flux within the
atmosphere and ocean are still sub-Eddington because the potential
energy of gravitational infall has been released long before material
reaches these depths (where it is now in hydrostatic balance). We
sometimes consider such super-Eddington rates because they are required
for the thermal stability of the accreting envelope.

  The continuity equation for an element $i$ with mass fraction
$X_i=\rho_i/\rho$ is
\begin{eqnarray}
        \frac{\partial X_i}{\partial t}
        + \dot{m} \frac{\partial X_i}{\partial y}
                = \frac{A_im_p}{\rho}( -r_i^{dest}+r_i^{prod}),
\end{eqnarray}
where $r_i^{dest}$ ($r_i^{prod}$) is the particle destruction
(production) rate, $A_i$ is the baryon number of species $i$,
and $v=-\dot{m}/\rho$ is the downward flow velocity
(Wallace, Woosley \& Weaver 1982; Bildsten, Salpeter \& Wasserman 1993).
We assume steady state burning and set $\partial X_i/\partial t = 0$.
For the \ta reactions we use the energy generation rate from
Fushiki \& Lamb (1987a). At larger depths helium is burned in alpha
captures to form heavier elements. For the
$^{12}\textrm{C}(\alpha,\gamma)^{16}\textrm{O}$ reaction we use the
reaction rate from Buchmann (1996, 1997) with the fitting formula
that gives $S=146$ keV barns at 300 keV center-of-mass energy.
Caughlan \& Fowler (1988) is used for all further reactions which
includes $^{16}\textrm{O}(\alpha,\gamma)^{20}\textrm{Ne}$,
$^{20}\textrm{Ne}(\alpha,\gamma)^{24}\textrm{Mg}$,
and $^{24}\textrm{Mg}(\alpha,\gamma)^{28}\textrm{Si}$. Possible subsequent
burning of silicon does not occur for the accretion rates we consider.
We follow Schalbrock et al. (1983) and set the uncertainty factor
in the $^{20}\textrm{Ne}(\alpha,\gamma)^{24}\textrm{Mg}$ and
$^{24}\textrm{Mg}(\alpha,\gamma)^{28}\textrm{Si}$ reaction rates to 0.1,
and screening is included following Salpeter \& Van Horn (1969).
\newcounter{subequation}[equation]
\renewcommand{\theequation}{\arabic{equation}\alph{subequation}}
%

  Since helium rich accreting material is most likely from a helium
white dwarf companion of the NS, a more realistic model should include
a fraction of $^{14}$N in the composition. This is the most abundant
heavy nucleus in the accreting material as this nucleus is the pile-up
point during the catalytic CNO cycle in the progenitor of the helium
white dwarf. Its abundance could be as large as 2\%, and the only
available fusion reaction in the accreting NS envelope is
$^{14}{\rm N}(\alpha,\gamma)^{18}{\rm F}$, which releases 4.415 MeV.
Using the rates from Caughlan \& Fowler (1988)
(also see Couch et al. 1972), we integrated the continuity
equation for $^{14}$N, finding that this element is quickly converted to
$^{18}$F at a shallow depth of $10^7 \textrm{ g cm}^{-2}$.
Unless another $\alpha$ capture occurs, the $^{18}$F
will decay by positron emission to $^{18}$O with a half-life of
\mbox{109.8 minutes}, creating a slight gradient in $\mu_e$
(the mean molecular weight per electron) in the envelope. By the
time this decay occurs, all the $\alpha$ particles will have been
consumed so that the usual closure of this chain to $^{22}$Ne seems
unlikely. This consumption of $\alpha$ particles is not dramatic enough
to drastically affect the composition of other heavy $\alpha$-capture
elements, nor affect the temperature sensitivity of \ta reactions which
leads to instability of a mode. We therefore omit $^{14}$N from our
calculations. It is interesting to note that we do
find a range of high accretion rates ($\dot{m}\sim 5 \dot{m}_{\rm Edd}$) where
a thermal instability in the $^{14}{\rm N}(\alpha,\gamma)^{18}{\rm F}$
reaction will occur before \ta in an accumulating envelope. The small
amount of energy released in such unstable reactions would have little
observational effect. An accreting pulsar such 4U 1626--67 might have the
correct conditions for such a scenario to occur (if there is
super-Eddington accretion occuring at its polar cap), but the flares that
would be created by such unstable reactions (which would occur every
$\sim100$ s) would never exceed $\sim10^{-3}$ of the accretion
luminosity.

  The entropy equation is
\begin{eqnarray}
	T\frac{ds}{dt}
	= - \frac{1}{\rho} \vec{\nabla} \cdot \vec{F} + \epsilon,
\end{eqnarray}
where $s$ is the specific entropy (which has units of erg g$^{-1}$ K$^{-1}$),
$\epsilon$ is the energy generation rate from nuclear reactions, and the
flux leaving a plane-parallel atmosphere is given by the standard
radiative transfer equation,
\begin{eqnarray}
        F = \frac{4acT^3}{3\kappa}\frac{dT}{dy},
\end{eqnarray}
where $\kappa$ is the opacity, and $a$ is the radiation constant.
Rewriting the specifc entropy as a function of $T(y,t)$ and $P(y,t)$,
equation (3) becomes
\begin{eqnarray}
	\frac{\partial F}{\partial y} + \epsilon
	= c_p \left( \frac{\partial T}{\partial t}
		+ \dot{m} \frac{\partial T}{\partial y} \right)
	- \frac{c_pT\dot{m}}{y}\nabla_{ad},
\end{eqnarray}
where $c_p$ is the specific heat at constant pressure and
$\nabla_{ad} \equiv (\partial \ln T/\partial \ln P)_s$.
In steady state $\partial T/\partial t = 0$ so that
\begin{eqnarray}
	\frac{dF}{dy} =  -\epsilon
		+ c_p\dot{m}\left( \frac{\partial T}{\partial y}
				-\frac{T}{y}\nabla_{ad}\right),
\end{eqnarray}
where the last term is often referred to as gravitational compression.

  Equations (2), (4), and (6) can now be solved numerically to find the
structure of the envelope. We relate pressure and density using the analytic
equations of state by Paczy\`{n}ski (1983b). The opacity in equation (4) is
set by electron scattering, free-free, and conductive opacities. We use the
electron scattering opacity as approximated in Paczy\`{n}ski (1983b).
The free-free opacity is from Clayton (1983),
\begin{eqnarray}
	\kappa_{\rm ff} = 0.753 \textrm{ cm}^2 \textrm{ g}^{-1}
		\frac{\rho_5}{T_8^{7/2}}
		\sum_i
		\frac{Z_i^2X_i}{A_i}
		g_{\rm ff}(Z_i,T,n_e),
\end{eqnarray}
where the sum is over all nuclear species,
$\rho_5 = \rho/(10^5 \textrm{ g cm}^{-3}$),
$T_8 = T/(10^8 \textrm{ K})$, $Z_i$ is the charge of the nucleus
of species $i$, and $g_{\rm ff}$ is the dimensionless Gaunt factor
that is analytically fit in Schatz et al. (1999). This factor takes
into account Coulomb wavefunction corrections, degeneracy, and
relativistic effects. Free-free is never the dominant opacity, since
when it becomes large the heat transport is mediated by conduction,
but it must be included to correctly calculate the total opacity
in the deeper regions of the envelope. The conductive opacity is
from Schatz et al. (1999) which uses the basic form of
Yakovlev \& Urpin (1980).

  Solving these equations requires setting boundary conditions for
the flux and temperature. The flux exiting the envelope is the sum
of the steady state nuclear burning flux, the internal gravitational
energy release, and any additional flux coming from the crust,
$F_{\rm crust}$. The radiative zero nature of the outer boundary means
that the surface temperature boundary condition is fairly simple
(Schwarzschild 1958). Near the top of the envelope, where the flux is
constant, hydrostatic balance can be combined with radiative diffusion,
equation (4), to find $dP/dT\propto T^3/\kappa$.
Integrating this result with the assumption that the opacity
is nearly constant (electron scattering is the dominate opacity)
provides the relation $P\propto T^4$ plus a constant
of integration set by the initial value of the temperature.
Since $T^4\gg T_{\rm eff}^4$ this constant is dwarfed at the
depths of interest ($y \gtrsim 10^6\textrm{ g cm}^{-2}$)
making the initital choice of temperature negligible.
Figure 1 shows nuclear compositions of envelopes
as a function of column depth for different accretion rates and
$F_{\rm crust}=0$. The helium mass fraction is depleted to
$\approx50\%$ near a column density of $5\times10^7 \textrm{ g cm}^{-2}$ (the
burning layer). At high accretion rates, considerable amounts of heavy elements
are formed in the deep ocean. Figure 2 shows the flux and temperature profiles
for these same accretion rates. As the accretion rate increases, the
temperature profile becomes hotter.
Since these envelopes have no flux from the neutron star crust,
the temperature is approximately isothermal (except from gravitational
settling contributions) below the burning depth. In Figure 3,
$F_{\rm crust}$ is increased and the temperature profile becomes
hotter and non-isothermal at large depths. In \S 3.3 we show how this
extra flux alters the conditions for mode stability.

  An important quantity when solving for the modes
is the Brunt-V\"{a}is\"{a}l\"{a} frequency (Bildsten \& Cumming 1998),
\begin{eqnarray}
	N^2 &=& \frac{g}{h} \left\{ \frac{\chi_T}{\chi_\rho}
			\left[ \nabla_{ad}
				- \left( \frac{d\ln T}{d\ln P} \right)_\ast
			\right]
		\right.
		\nonumber \\
		&&\left. { }- \frac{\chi_{\mu_e}}{\chi_\rho}
			\left( \frac{d\ln\mu_e}{d\ln P} \right)_\ast
		- \frac{\chi_{\mu_i}}{\chi_\rho}
			\left( \frac{d\ln\mu_i}{d\ln P} \right)_\ast\right\},
\end{eqnarray}
where $\chi_Q = \partial\ln P/\partial\ln Q$ with all other intensive
variables set constant, and $\ast$ refers to derivatives of the
envelope's profile. Figures 2 and 3 shows $N$ versus accretion rate
and base flux. The general profile of $N$ is set by the increase in
the scale height with depth because $N^2\sim g/h$, but there is also
a bump near the helium burning depth. This added buoyancy is caused
by the change in the mean molecular weight as helium burns to carbon,
and it has important consequences for the modes. At high
accretion rates $N$ shows additional structure due to the heavier
elements that are being made, but this has little effect on the
modes.

\subsection{Applicability of Steady State Burning Models}

  It is important to understand when a steady state burning
envelope will be thermally stable. One way is to have an
envelope hot enough that the temperature sensitivity
of \ta reactions is weaker than that of cooling. This is
easily accomplished at high accretion rates. A one-zone
model approximates the competition between nuclear heating
and radiative cooling in the helium burning layer. This
argument (Fujimoto, Hanawa \& Miyaji 1981) is reviewed in
Bildsten (1998) and repeated here for completeness. For \ta
reactions we write $\epsilon_{3\alpha}\propto \rho^\eta T^\nu$.
The cooling is set by the derivative of radiative transfer,
equation (4), which in the one-zone approximation becomes
\begin{eqnarray}
	\frac{\partial F}{\partial y} \approx - \frac{acg^2T^4}{3\kappa P^2}
	\equiv \epsilon_{\rm cool}.
\end{eqnarray}
Comparing the temperature sensitivity of $\epsilon_{3\alpha}$ and
$\epsilon_{\rm cool}$ gives a local condition for stability as
$d\epsilon_{3\alpha}/dT<-d\epsilon_{\rm cool}/dT$. This can be written
more explicitly in the steady state case
($\epsilon_{3\alpha}=-\epsilon_{\rm cool}$) as
\begin{eqnarray}
	\nu-4+\frac{d\ln\kappa}{d\ln T}+\frac{d\ln\rho}{d\ln T}
		\left( \eta + \frac{d\ln\kappa}{d\ln\rho} \right) < 0.
\end{eqnarray}
Approximating $\nu = -3 +44/T_8$, $\eta=2$, $d\ln\kappa/d\ln T = 0$,
$d\ln\kappa/d\ln\rho = 0$, and $d\ln\rho/d\ln T = -1$, we find
$T>4.8\times10^8\textrm{ K}$ is needed for stable nuclear burning
of pure helium. We then look at Figure 2 to find the corresponding
accretion rate. From this we find that accretion must be at least $5-10$
times the Eddington rate for stable burning.

  Another way to suppress thermal instabilities is
to have a large \fc (Paczy\'{n}ski 1983a; Bildsten 1995).
There is suggestive observational evidence of a lack of
type I X-ray bursts in the days following a superburst
(Cornelisse et al. 2000; Kuulkers et al. 2002)
which may be due to the additional flux from cooling carbon ashes
(as speculated in Cumming \& Bildsten 2001). For these reasons
we also study the mode structure for envelopes with high \fc.


\section{Nonradial Oscillations in a Thin Shell}

  Nonradial oscillations have been studied in detail for NSs both
in isolation (McDermott, Van Horn \& Scholl 1983;
Finn 1987; McDermott, Van Horn \& Hansen 1988; McDermott 1990;
Reisenegger \& Goldreich 1992; Strohmayer 1993) and accreting
(McDermott \& Taam 1987; Bildsten \& Cutler 1995, hereafter BC95;
Bildsten, Ushomirsky \& Cutler 1996, hereafter BUC96;
Strohmayer \& Lee 1996; Bildsten \& Cumming 1998).
To solve the nonradial perturbations on
a rotating neutron star we follow BUC96,
but we also include additional new strategies that
assist in solving these equations.

\subsection{Adiabatic Perturbation Equations}

  In the outer envelope (at $y\approx10^6\textrm{ g cm}^{-2}$) the thermal
timescale is $t_{\rm th}\approx y c_p T/F \approx 0.1 \textrm{ sec}$.
This will be shown to be shorter than the g-mode periods.
This means we can consider purely adiabatic perturbations of
as long as we confine our analysis to large enough depths
($y\gtrsim10^6 \textrm{ g cm}^{-2}$). Damping will occur in
those regions where the mode period is longer than the
thermal time.

  For the perturbations, we consider conservation of mass,
$\partial \rho/\partial t+ \vec{\nabla} \cdot ( \rho \vec{v} ) = 0$,
and momentum,
\begin{eqnarray}
	\rho \left( \frac{\partial}{\partial t}
		+ \vec{v} \cdot \vec{\nabla} \right) \vec{v}
	= - \vec{\nabla} P - \rho \vec{\nabla} \Phi
	- 2 \vec{\Omega} \times \vec{v},
\end{eqnarray}
where $\Phi$ is the gravitational potential, $\vec{\Omega}$ is the
neutron star spin vector, and the last term is the Coriolis force.
The centrifugal force can be neglected because
$\Omega \ll (GM/R^3)^{1/2}$ for most accreting NSs
(Chakrabarty et al. 2003). We make Eulerian
perturbations of these conservation equations, substituting for the
dependent variables $Q \rightarrow Q_0 + \delta Q$, where $Q_0$ denotes
the static background (this subscript is dropped in subsequent
expressions).  Perturbations are assumed to have the form
$\delta Q= \delta Q(r,\theta)\exp(im\phi + i\omega t)$ where $\omega$
is the mode frequency in the rotating frame (an observer will see
a frequency $\omega_{\rm obs} = |\omega-m\Omega|$). Since the
background model has no fluid motion we substitute
$\vec{v} \rightarrow \delta \vec{v} = d\vec{\xi}/dt$,
where $\vec{\xi}$ is the local Lagrangian displacement.
Keeping terms of linear order, continuity becomes
\begin{eqnarray}
	\delta \rho + \vec{\nabla} \cdot ( \rho \vec{\xi} ) = 0,
\end{eqnarray}
and the momentum equation can be broken into its three components to
give
\begin{eqnarray}
        \addtocounter{subequation}{+1}
	- \rho \omega^2 \xi_r
        	&=& - \frac{\partial\delta P}{\partial r}
		- g \delta \rho
        	+ 2 i \Omega \rho \omega \xi_\phi \sin\theta,
	\\
	\addtocounter{equation}{-1}
        \addtocounter{subequation}{+2}
	- \rho \omega^2 \xi_\theta
                &=& - \frac{1}{R} \frac{\partial\delta P}{\partial \theta}
                + 2 i \Omega \rho \omega \xi_\phi \cos\theta,
	\\
	\addtocounter{equation}{-1}
        \addtocounter{subequation}{+3}
	- \rho \omega^2 \xi_\phi
                &=& - \frac{1}{R\sin\theta}
			\frac{\partial\delta P}{\partial \phi}
                - 2 i \Omega \rho \omega 
			(\xi_r \sin\theta + \xi_\theta \cos\theta).
\end{eqnarray}
These equations can be further simplified using the 
``traditional approximation''
(Chapman \& Lindzen 1970; Brekhovskikh \& Goncharov 1994; BUC96),
allowing for separable solutions. The first simplification is
neglecting the Coriolis force in equation (13a) for
$\Omega \ll N^2h/\omega R \sim 10^5 \textrm{ Hz}$. In the
non-rotating case incompressibility tells us that
$\xi_r/\xi_\theta \ll 1$ so we can also neglect the $\xi_r$ term
in equation (13c). Essentially we are assuming that the Coriolis
force has no effect in the radial direction because of the strong
gravitational field present and the thin size of the envelope. In
the adiabatic limit $\Delta P/P = \Delta \rho/(\Gamma_1 \rho)$, where
$\Delta$ is a Lagrangian perturbation
($\Delta Q = \delta Q + \vec{\xi} \cdot \vec{\nabla}Q$) and
$\Gamma_1$ is the adiabatic exponent,
$\Gamma_1 \equiv (\partial\ln P/\partial\ln\rho)_{s}$.
The radial equation then becomes identical to the non-rotating case,
\begin{eqnarray}
        \frac{d}{dr} \frac{\delta P}{P}
        = \left( 1 - \frac{1}{\Gamma_1}  \right) \frac{1}{h} \frac{\delta P}{P}
        + \left( \frac{\omega^2}{g} - \frac{N^2}{g} \right) \frac{\xi_r}{h}.
\end{eqnarray}
Combining equation (13b) with continuity, equation (12), results in
\begin{eqnarray}
        \frac{d\xi_r}{dr}
        = \frac{\xi_r}{\Gamma_1h}
        - \frac{1}{\Gamma_1 } \frac{\delta P}{P}
        - \frac{gh}{\omega^2 R^2}
                L_\mu\left[ \frac{\delta P}{P} \right],
\end{eqnarray}
identical to the non-rotating result except now $-l(l+1)$ has been
replaced by $L_\mu$ which is an angular operator, independent of the
NS's composition, defined as
\begin{eqnarray}
	L_\mu &=& \frac{\partial}{\partial\mu}
		\left[ \frac{1-\mu^2}{1-q^2\mu^2}
			\frac{\partial}{\partial\mu}\right]
		\nonumber \\
		&&- \frac{m^2}{(1-q^2\mu^2)(1-\mu^2)}
		- \frac{qm(1+q^2\mu^2)}{(1-q^2\mu^2)^2},
\end{eqnarray}
where $q=2\Omega/\omega$ and $\mu=\cos\theta$. The angular piece is now
a separate eigenvalue equation,
\begin{eqnarray}
         L_\mu\left[ \frac{\delta P}{P} \right]
        = -\lambda \frac{\delta P}{P},
\end{eqnarray}
where $\lambda$ is the ``effective wavenumber'' (BUC96).

  We first focus on solving the radial eigenfunction equations in
\S3.2, and then return to solve for rotational modifications in
\S3.4 by solving for $\lambda$. In the limit that
$\omega^2\ll N^2$ (the limit for g-modes) it is easy to show that the
frequency scales like $\omega\propto l(l+1)^{1/2}$, and therefore
in the rotating case $\omega\propto\lambda^{1/2}$. We solve the
radial structure for $l=1$ with no rotation and define this
frequency as $\omega_0$, later scaling to the rotationally modified
frequency by setting $\omega = \omega_0(\lambda/2)^{1/2}$.

\subsection{Solving the Radial Mode Structure}

  We now use equations (14) and (15) to solve for the radial
eigenmode structure where we set $L_\mu[\delta P/P]=-l(l+1) \delta P/P$
and $l=1$. In this section and \S3.3, instead of considering
$\xi_r$, $\xi_\theta$, and $\xi_\phi$ we consider only $\xi_r$ and
$\xi_\perp$, where $\xi_\perp$ is the vector displacement field
perpendicular to $\xi_r$. We confine the radial structure to lie between
$y \approx 10^6 \textrm{ g cm}^{-2}$ and $10^{13} \textrm{ g cm}^{-2}$.
The top boundary (where we set $\Delta P=0$) is set by our adiabatic
approximation -- we need to be at a sufficient depth so that the
thermal time scale is longer than the mode period. In general this
will be different for each mode and each envelope model, but it
is always in the range of $10^5-10^6 \textrm{ g cm}^{-2}$
(see Appendix A of BC95 for how to deal with the transition to the
outer atmosphere). The bottom boundary (where $\xi_r=0$) is set at
the crust where the material becomes rigid. Fortunately, the properties
of the mode of most interest are nearly independent of this bottom
boundary condition.
 
  We numerically solve for the eigenfunctions, starting at the surface
and shooting for the bottom boundary. As examples, we plot the
first three g-modes with zero crossings (these will henceforth be
referred to as the $n=2,3,\textrm{ and }4$ g-modes) in Figure 4
for an accretion rate of $5.0 \dot{m}_{\rm Edd}$ and \fc$=0$. The
perpendicular displacement is given by
\begin{eqnarray}
	|\xi_\perp| = \sqrt{l(l+1)}
			\frac{g h}{R \omega^2}
			\left| \frac{\delta P}{P} \right|.
\end{eqnarray}
The normalization is arbitrarily set to
$\delta P/P =1$ at the surface, but the ratio in sizes of
$\xi_r$ to $\xi_\perp$ is meaningful and well approximated by
the incompressible limit (which predicts $\xi_r/\xi_\perp\sim h/R$).
Along with the radial and perpendicular displacements
we include the logarithmic kinetic energy density of the modes
\begin{eqnarray}
	\frac{dE}{d\ln y} = \frac{1}{2} 4 \pi R^2 \omega^2
		|\xi_\perp|^2 y,
\end{eqnarray}
where we have used the approximation $\xi_r/\xi_\perp\sim h/R\ll1$.
This quantity shows where the mode ``lives'' in the envelope.
The first mode in Figure 4 has a fairly constant $\xi_\perp$
above the burning layer, and its amplitude falls off abruptly
at larger depths. This mode, which we refer to as the ``shallow surface wave,''
resides primarily in the shallow helium layer above the carbon ash
ocean and is fairly insensitive to the bottom boundary condition.
This is also reflected by $dE/d\ln y$ which is small at the bottom of
the ocean.
On the other hand, the other modes in Figure 4, with a larger number
of nodes, have energy densities that are nearly uniform as the frequency
decreases. This shows that these modes will be sensitive to conditions
at all depths including the bottom boundary.

  The g-mode frequencies versus accretion rate are summarized in Figure 5.
These approximately scale $\propto \dot{m}^{1/8}$ as we would expect from
combining hydrostatics with radiative transfer
and assuming $\omega^2 \sim ghk^2$, where $k^2=l(l+1)/R^2$.
The modes' frequency dependence on \fc is shown in Figure 6.
The shallow surface wave ($n=2$) changes the least as \fc is changed
because its energy density is confined above the ocean. All of the $l=1$
frequencies are within the range of 10--30 Hz for a few nodes and decrease
for higher order g-modes, as expected from WKB analysis.

\subsection{Mode Stability}

  We now address the linear stability of these modes by considering the
exchange of energy between the mode and the star. This involves
calculating the ``work integral'' for each mode
(Cox 1980; Unno et al. 1989). In the adiabatic non-rotating case
that we considered in \S 3.1, equations (12) and (13) can be combined 
into one Hermitian operator, $\mathcal{L}$, acting on $\vec{\xi}$ which is
\begin{eqnarray}
        \mathcal{L}(\vec{\xi}) &=&
		\frac{1}{\rho^2} (\vec{\nabla}P)
			\vec{\nabla}\cdot(\rho\vec{\xi})
		\nonumber \\
		&& - \frac{1}{\rho}\vec{\nabla}(\vec{\xi}\cdot\vec{\nabla}P)
		- \frac{1}{\rho}\vec{\nabla}
			(\Gamma_1P\vec{\nabla}\cdot\vec{\xi}),
\end{eqnarray}
where
\begin{eqnarray} 
        \frac{d^2\vec{\xi}}{dt^2} = - \mathcal{L}(\vec{\xi})
                = - \omega^2 \vec{\xi}.
\end{eqnarray}
From this eigenvalue equation
the clear interpretation of $\mathcal{L}$ is that it gives the
force per mass for a given vector displacement field $\vec{\xi}$.
Since $\mathcal{L}$ is Hermitian, we know that $\omega^2$ must be purely
real and therefore there are only damped or oscillatory solutions.

  In the nonadiabatic case, the energy equation must also
be included along with continuity and momentum conservation. This additional
equation is written in the form
\begin{eqnarray}
	\frac{d\ln P}{dt}
		= \Gamma_1 \frac{d\ln \rho}{dt}
		+ (\Gamma_3-1) \frac{\rho T}{P} \frac{ds}{dt},
\end{eqnarray}
where $\Gamma_3$ is the adiabatic exponent,
$\Gamma_3-1 \equiv (\partial\ln T/\partial\ln\rho)_s$.
Taking perturbations and linearizing equation (22), followed
by combining it with equations (12) and (13), the resulting expression
can be manipulated to get
\begin{eqnarray}
	\frac{d}{dt} \left\{\int_V \frac{1}{2}\rho
			\left|\frac{d\vec{\xi}}{dt}\right|^2d\tau
			+ \int_V \frac{1}{2}\vec{\xi}
				\cdot{\mathcal L}(\vec{\xi})\rho d \tau\right\}
	\nonumber \\
	= - \int_V (\Gamma_3-1) T\Delta s \frac{d}{dt}
		\frac{\Delta\rho}{\rho} \rho d\tau.
\end{eqnarray}
The first term on the left is the kinetic energy of the mode while the
next term is the force dotted with the displacement, i.e. the work. We
therefore identify the left hand side as the change in energy per unit
time for the mode. The right hand side is then an equivalent expression
for the change in energy. Averaging over one mode period and integrating
by parts along with making the substitution
\begin{eqnarray}
	T\frac{d\Delta s}{dt} = \Delta
		\left(\epsilon - \frac{1}{\rho}\vec{\nabla}\cdot\vec{F}\right),
\end{eqnarray}
we obtain the the familiar work integral,
\begin{eqnarray}
	\left< \frac{d\Psi}{dt} \right>
	= \frac{1}{\Pi}\int_0^\Pi dt
	\int_V (\Gamma_3-1) \frac{\Delta\rho}{\rho}
	\Delta \left(\epsilon - \frac{1}{\rho}\vec{\nabla}\cdot\vec{F}\right)
	\rho d\tau,
\end{eqnarray}
where $\Psi$ is the mode energy, $\Pi$ is the mode period,
and $d\tau=4\pi R^2dr$ is a volume element. All
perturbations are evaluated using an adiabatic eigenfunction so that this
represents the first order imaginary correction to the mode frequency.
If this integral is positive (negative) the mode will be excited (damped).
It is important to note that just as for the thin shell thermonuclear
instability, mode stability depends on a competition between heating
($\epsilon$) and cooling ($\vec{\nabla}\cdot\vec{F}/\rho$). However,
in the case of a mode, this difference is integrated over the entire
envelope, weighted by the mode's eigenfunction,  giving a mathematically
different stability criteria.

  Figure 7 shows the logarithmic integrand from equation (25),
\begin{eqnarray}
	\textrm{Heating vs. Cooling}
	= (\Gamma_3-1) \frac{\Delta\rho}{\rho}
        \Delta
	\left(\epsilon - \frac{1}{\rho}\vec{\nabla}\cdot\vec{F}\right)
	y,
\end{eqnarray}
for three different modes at the same accretion rate. Regions where this
function is positive (negative) show where the mode is being
pumped (damped). It is clear that
the $n=2$ mode is excited while the others are damped. This excitation
comes from an increase in nuclear reactions at the burning depth
($y\approx5\times10^7\textrm{ g cm}^{-2}$).
By dividing the work integral by the integral of the energy density,
equation (19), we derive an excitation rate,
\begin{eqnarray}
	\kappa_{\rm excite}
		= \left< \frac{d\Psi}{dt} \right>
		  \left[ \int\frac{dE}{d\ln y}d\ln y \right]^{-1}.
\end{eqnarray}
Figures 8 and 9 show this rate as a function of $\dot{m}$  and
\fc and show that the excitation timescale can be as fast as
less than a second for
the shallow surface wave. All higher order g-modes are always damped
($\kappa_{\rm excite}<0$).

  The work integrals calculated in this section are valid for all of the
rotationally modifed modes we consider in the next section, including
the Kelvin modes and the r-modes. In the ``traditional approximation''
(\S 3.1) all of the inertial modes have similar radial eigenfunctions because
buoyancy dominates over the Coriolis force. The work integral only
depends on the radial eigenfunction so that it will have the same valule
for each of the modes. The only differences between the rotationally
modified modes' stabilities are their excitation rates.
These depend on their kinetic energies, which are
related to their frequencies, see equation (19).
These frequencies are altered due
to the Coriolis force as explained in the following section.

\subsection{Solving the Angular Equation}

  We now discuss our methods for solving equation (17) that
build on BUC96. This provides us with eigenfunctions that span the
latitudinal length of the star and relates $\lambda$ to $q$.
Equation (17) is known as Lapace's tidal equation, and its solutions are
Hough functions. For a standard method of how to solve this equation as
an expansion of spherical harmonics for $q\sim1$ see Longuet-Higgins (1968).
However, since we are interested in large values of $q$ (for which modes
can decay exponentially towards the pole, requiring a large number of terms)
we instead follow BUC96 and numerically solve this equation. The
numerical challenges are singularities at $\mu=1$ and $\mu=1/q$ which
were dealt with first in BUC96. An easier way to get around the
singularity at $1/q$ is to break equation (17) into two first order
differential equations.\footnotemark We define $a\equiv\delta P/P$ and
\begin{eqnarray}
        b \equiv - \frac{1-\mu^2}{1-q^2\mu^2}
                \left[ \frac{\partial a}{\partial\mu}
                        - \frac{qm\mu}{1-\mu^2} a \right].
\end{eqnarray}
Essentially we are setting $b\propto\sin\theta\xi_\theta$.
Substituting these relations into Laplace's tidal equation results in
two first order differential equations
\begin{eqnarray}
        (1-\mu^2) \frac{\partial a}{\partial\mu}
                = qm\mu a - (1-q^2\mu^2) b
        \\
        \frac{\partial b}{\partial\mu}
                = \left(\lambda - \frac{m^2}{1-\mu^2} \right) a
                        - \frac{qm\mu}{1-\mu^2} b.
\end{eqnarray}
These equations can be further simplified by factoring out the singularity
at $\mu=1$. This is done by defining $a\equiv (1-\mu^2)^{|m|/2}A$
and $b\equiv (1-\mu^2)^{|m|/2}B$ which provides the equations
\begin{eqnarray}
        (1-\mu^2) \frac{\partial A}{\partial\mu}
                = (|m|+qm)\mu A - (1-q^2\mu^2) B
        \\
        \frac{\partial B}{\partial\mu}
                = \left(\lambda - \frac{m^2}{1-\mu^2} \right) A
                        + \frac{(|m|-qm)\mu}{1-\mu^2} B.
\end{eqnarray}
\footnotetext{We thank Phil Arras and Greg Ushomirsky for pointing out this
simplification.}
In this new form the angular eigenvalue equation is fairly easy to solve.

  We begin integration at $\mu=1-\epsilon$ (where $\epsilon$ is some small
positive number), with the conditions that $A=1$ and
$B=(|m|+qm)\mu/(1-q^2\mu^2)$ (the normalization of $A$ is arbitrary
because the equations are linear). The symmetry of the solutions
allows us to shoot numerically toward $\mu=0$, where the boundary
condition is $A(0)=0$ (for odd solutions) or $A'(0)=0$ (for even solutions).
The eigenvalues are tracked by starting at small $q$ (little
rotation) where $\lambda = l(l+1)$ and then increasing $q$ while solving
for the boundary condition at the equator.

  Figure 10 shows the eigenvalues of the rotationally modified
g-modes that correspond to $l=1$ or $l=2$ in the non-rotating
limit as a function of $q$ (as can be seen because
$\lambda = 2$ or 6 for $q\ll1$) (consistent with BUC96).
In all cases, the lowest $m$ value mode
for a given $l$ has the interesting feature that at large $q$ it asymptotes
to $\lambda=m^2$. This is the ``Kelvin mode'' (Longuet-Higgins 1968).
All other g-modes increase in $\lambda$ for increasing $q$ and asymptote
to $\lambda\propto q^2$ (as discussed in Papaloizou \& Pringle 1978 and BUC96).

  Since the ``traditional approximation'' captures all rotationally
modifed modes of the star, we have also included in Figure 10 two of
the odd parity r-modes (not shown in BUC96). These have the property
that at small $q$ there is a simple relationship between the neutron
star spin and the mode frequency, namely $\omega = 2 m\Omega/l(l+1)$
(Saio 1982; Lee \& Saio 1986). At higher spin frequencies, the mode
frequencies are shown to deviate from this relationship.
Additional even and odd parity r-modes are shown in Figure 11.
The r-modes all have
$m>0$ so they will move retrograde with respect to the neutron star spin.
They would therefore appear below the spin frequency in observations
(but not with the same spacing as the Kelvin mode).

  Figure 12 shows some examples of the eigenfunctions that are found to
span the latitudinal direction. It is important to note that for the
rotationally modified g-modes, the eigenfunction becomes concentrated
at the equator for higher spin rates. The modes are exponentially
damped for $\mu\gtrsim1/q$. On the other hand, other modes, such as the
Kelvin mode, are not as affected by rotation and will have a larger
amplitude covering a larger area of the surface.

\section{Observational Tests}

  The closest physical realization of the model we describe in the previous
sections is 4U 1820--30. This LMXB sits near the center of the globular
cluster NGC 6624 and was the first to show type I X-ray bursts
(Grindlay et al. 1976). It is also the LMXB with the shortest
orbital period of 11.4 minutes (Stella, White \& Priedhorsky 1987).
Since the system is so compact, the donor star is most likely a
$0.06-0.08 M_\odot$ helium white dwarf (Rappaport et al. 1987) which means
that the accreted material is very helium rich. This is consistent with
the cooling timescale of X-ray bursts (10--20 seconds) and the photospheric
radius expansions seen from this source
(for examples, see Haberl et al. 1987).

  4U 1820--30 has a $171.033\pm0.326$ day modulation between high and low
luminosity states (Priedhorsky \& Terrell 1984; Chou \& Grindlay 2001).
X-ray bursts are only seen during the low-luminosity state (Clark et al. 1977;
Stella, Kahn \& Grindlay 1984) which suggests that these modulations are
related to the accretion rate. Such a scenario is qualitatively consistent
with our understanding of thermal instabilities as discussed in \S 2.2. During
the low state the bursting properties, including fluences and recurrence
times, are well understood using current bursting models (Cumming 2003).
An outstanding problem still remains in understanding the high-luminosity
state because bursting is suppressed even though 4U 1820--30 is
still well below the Eddington rate. This might be related to a change in
the physics of the bursts from being convective to radiative in the
high state (Bildsten 1995). 4U 1820--30 has also shown a $\sim$3 hour
long thermonuclear burst (Strohmayer \& Brown 2001), or superburst, which is
likely from unstable carbon burning. Other LMXBs such as
4U 1735-44 and GX 17+2 show that type I bursting ceases for many days
following a superburst (Cornelisse et al. 2000;
Kuulkers et al. 2002) which has been hypothesized to be due to additional
flux due to cooling carbon ashes (Cumming \& Bildsten 2001).

  As was shown in \S 3.4, the mode frequencies can become highly modified
for the high spin rates (200--700 Hz) inferred from accreting X-ray pulsars
and X-ray burst oscillations (Chakrabarty et al. 2003). However,
4U 1820--30 has never shown burst oscillations and therefore has an
unknown spin period. Thankfully, we only have one unstable mode to consider,
so we show how its frequency is modified as a function of the spin
rate in Figure 13. By measuring multiple frequencies from 4U 1820--30
during one of its non-bursting states one could hope to constrain its
spin period. In Figure 14 we show a similar plot focusing on the most
probable spin rates for 4U 1820--30.

  The lack of pulsations from most LMXBs implies a low magnetic field.
This is helpful for the mode calculations as a large magnetic field
would resist shearing from the change in $\xi_\perp$ and modify the
mode frequency. Following BC95, we estimate the maximum magnetic field
before the shallow surface mode would be dynamically effected as
$B_{\rm dyn}^2 \approx 8 \pi h^2 \rho \omega^2$, resulting in
\begin{eqnarray}
	B_{\rm dyn} \approx 5 \times 10^7 \textrm{ G}
		\left( \frac{ \omega/(2\pi)}{21.4 \textrm{ Hz}}\right).
\end{eqnarray}
For the case of Kelvin or r-modes $B_{\rm dyn}$ is constrained to be
even smaller because these modes have smaller frequencies in the rotating
frame.

  Unfortunately, neither 4U 1820--30, nor any other helium accreting LMXBs
(such as the recently detected ultracompacts in outburst though they have
even lower accretion rates), has ever shown oscillations like the ones
we describe. This may not be surprising since during high accretion states
one might expect the surface of the neutron star to be obscured by
material that cannot make its way to the surface due to radiation pressure.
Furthermore, the rotationally modified eigenfunctions are squeezed toward
the equator (as shown in Figure 12) meaning that modes may be covered by
the accretion disk. We hope that by constraining the possible frequencies
at which modes could occur, and noting that mode excitation is best when
burning is steady, that searches for the modes will now be easier.

  This study also brings to light the importance that a shallow surface
wave might play for NSs. There are many other NS
envelopes that have the properties that this mode needs to exist --
namely a density or mean molecular weight discontinuity. In the short
time ($\sim$seconds) after a type I X-ray burst, the NS envelope is
composed of a hot, low density, layer of hydrogen and helium sitting
on top of a high density layer of heavy nuclei ashes. In this case the
discontinuity is even more extreme than the case we consider
because of the presence of hydrogen and the heavy
rapid proton capture elements in the ocean
(Wallace \& Woosely 1981; Hanawa, Sugimoto \& Hashimoto 1983;
Wallace \& Woosely 1984; Schatz et al. 1997, 1998; Koike et al. 1999).
Furthermore, if ignition of the type I X-ray burst were non-axisymmetric
it could drive modes such as these. Even though we have not evaluated
the stability, it seems likely that a shallow surface wave would
exist in such circumstances.  Previous theoretical studies predict
static frequencies (for example McDermott \& Taam 1987;
Strohmayer \& Lee 1996), but due to the temperature dependence of the
shallow surface wave ($\omega\propto T^{1/2}$) it would be shifting in
frequency and possibly not seen easily in a standard Fourier analysis.
A frequency scan (with changing $\omega$) would likely make these modes
easier to find. We will discuss the sign and sizes of these frequency
shifts, related to the NS temperature and spin, in a future paper.

\section{Conclusions and Discussion}

  This study makes clear the distinction between thermal and mode
perturbations of the thin envelope of a NS. We find one excited
mode on a thermally stable helium accreting NS which is a shallow surface
wave riding in the helium atmosphere above the ash ocean. It has a
frequency of $(20-30\textrm{ Hz})(l(l+1)/2)^{1/2}$ (depending on the
accretion rate) and has an excitation time of about a second. This mode is
modified by rotation to give multiple possible observable frequencies.
Seeing such a frequency pattern from a NS could be used to constrain
properties of the NS star (e.g. its spin or radius).
The rotationally modified eigenfunctions will have different levels of
visibility depending on both their final amplitudes and their
angular eigenvalues. The LMXB 4U 1820--30 is accreting helium rich material
and is the best candidate for having an envelope like the ones we consider.
No such oscillations have yet been reported from it, but perhaps now that
predictions have been made, a search for modes would be easier, especially
when bursting is absent. Making comparisons with just this
one system is limited, and it would be useful to do a
similar mode stability analysis on NSs that are accreting a solar mix
of hydrogen and helium, especially during times following a superburst
when type I bursts are not occuring.


  A further study could also address the eventual nonlinear evolution
of the excited mode. Such a study might predict observational
effects that this mode may exhibit. If this mode were to saturate
at a low amplitude (as is common in most other stars) then the most
observable feature would be the periodic modulation of the surface
luminosity. If the mode saturation is the result of mode coupling, then
other frequencies might also be visible. If the amplitude becomes
exceedingly nonlinear, the resulting large temperature excursions may well
lead to a more rapid consumption of the fuel than the rate of supply,
possibly triggering limit cycle behavior.

  This study highlights the importance a shallow surface wave may play
on NSs, and we plan to continue an investigation to understand better
what other NS environments might show this mode. An interesting study
would be the time before or after type I X-ray bursts when the NS envelope
should have the mean molecular weight discontinuity necessary to exhibit this
mode. In the latter case the frequency of the mode would be changing due
to the cooling of the atmosphere. This implies that shifting frequency
searches may be the correct way to find these modes. A future paper will
consider such effects in more detail, including the size and sign of the
expected shifts.

  Above all, this study has shown that spherically symmetric studies are
not complete and can easily miss important physics that may be relevant for
correctly understanding the properties of NSs. We hope to extend our work
to type I X-ray burst ignition, which has not been fully investigated in
the non-axisymmetric case. It is still somewhat unsatisfactory to think that
a NS ignites spherically all
at once, and our study makes clear that nonradial perturbations may give
different stability criterion for ignition than what has commonly been used.

  We thank Andrew Cumming for some of the original inspiration
for beginning this research and Phil Arras for discussions concerning
g-modes and rotational modifications. We have benefited greatly from
conversations with Philip Chang, Chris Deloye, Aristotle Socrates,
Anatoly Spitkovsky, Ronald Taam, Dean Townsley, and Greg Ushomirsky.
We thank the anonymous referee for suggestions which have added
clarity to this manuscript.
This work was supported by the National Science Foundation under
grants PHY99-07949 and AST02-05956 and by NASA through grant NAG
5-8658.  L. B. is a Cottrell Scholar of the Research Corporation.

\begin{figure}
\plotone{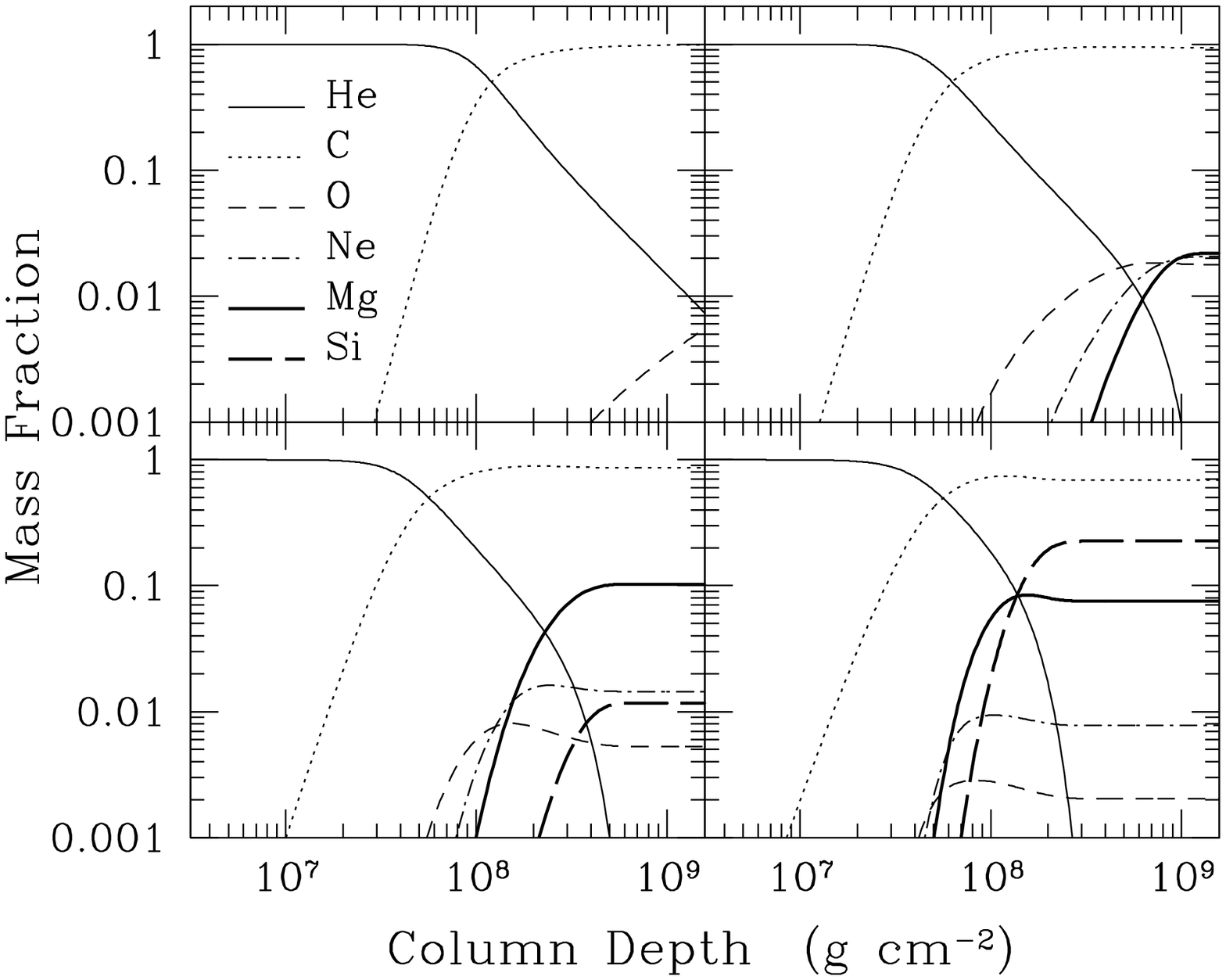} 
\figcaption{The composition of a neutron star envelope undergoing
pure helium accretion and burning in steady state.
The accretion rates are 1.0, 5.0, 10.0, and 20.0 (in units of
$\dot{m}_{\rm Edd}=1.5\times10^5\textrm{ g cm}^{-2}\textrm{ s}^{-1}$), starting
at the upper left and going clockwise, all with $F_{\rm crust}=0$.}
\end{figure}

\begin{figure}
\plotone{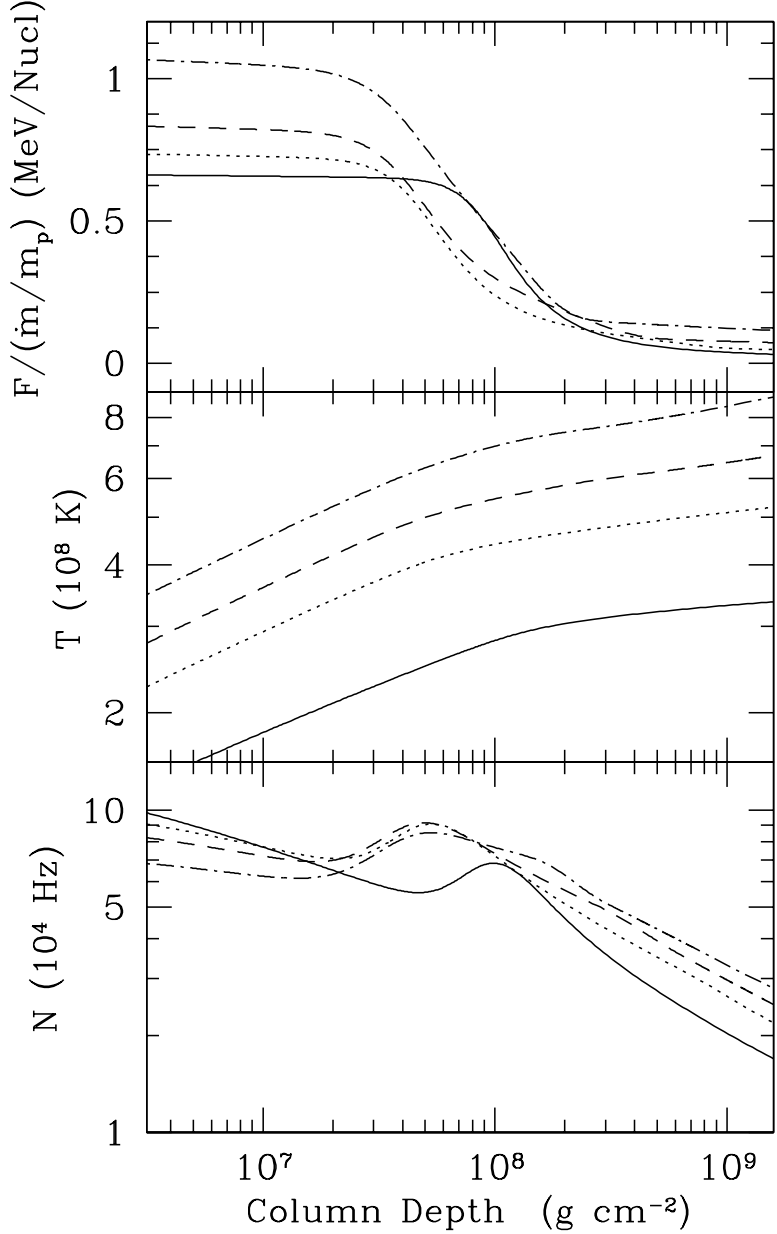}
\figcaption{The flux per accreted nucleon (in units of MeV per nucleon),
temperature, and Brunt-V\"{a}is\"{a}l\"{a} frequency, $N$, for
a neutron star undergoing pure helium accretion and
burning in steady state. This figure shows accretion rates of 1.0 (solid line),
5.0 (dotted line), 10.0 (dashed line), and 20.0 (dot-dashed line)
(in units of $\dot{m}_{\rm Edd}$), all with $F_{\rm crust}=0$
(at a column density of $10^{13} \textrm{ g cm}^{-2}$).
The Brunt-V\"{a}is\"{a}l\"{a} frequency shows a slight bump near
the burning layer due to additional buoyancy provided by
the change in mean molecular weight as helium burns to carbon.}
\end{figure}

\begin{figure}
\plotone{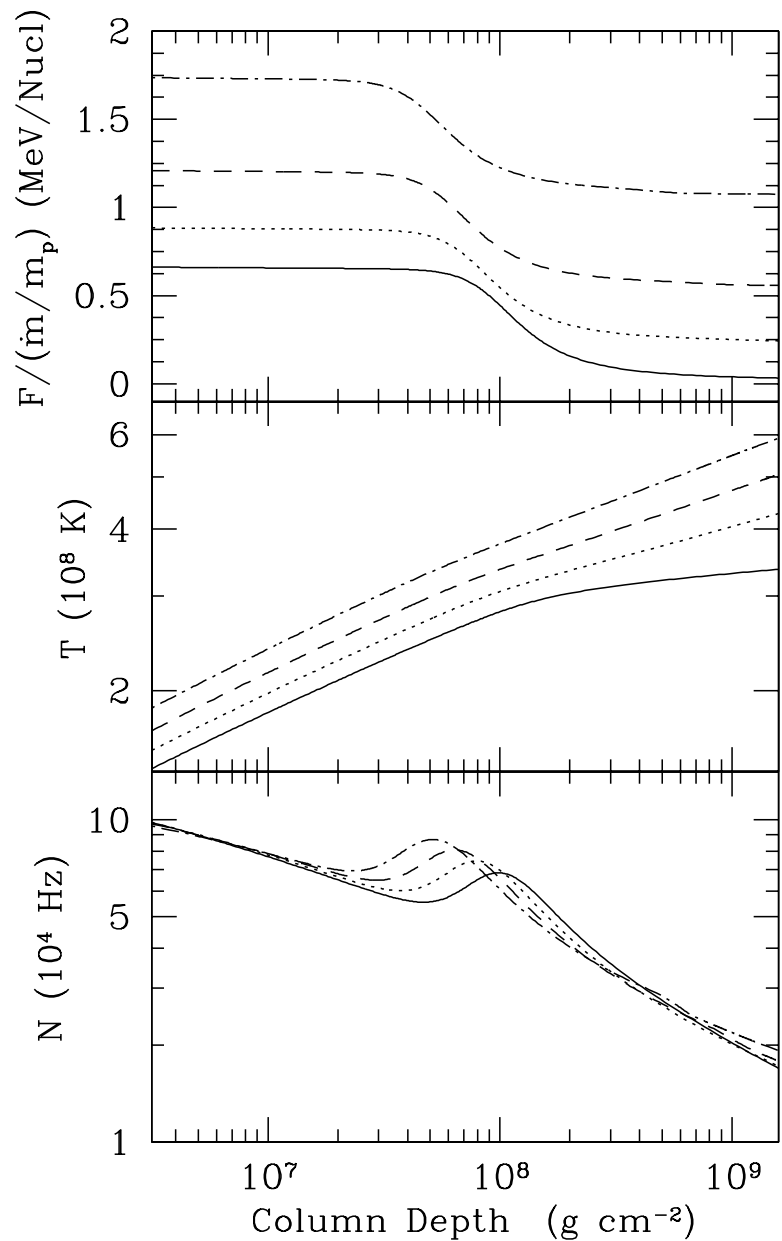}
\figcaption{Same as Figure 2, but with
$F_{\rm crust}$ values of 0.0 (solid line), 0.2 (dotted line), 0.5
(dashed line), and 1.0 (dot-dashed line)
(in units of MeV per accreted nucleon), all with
$\dot{m}=\dot{m}_{\rm Edd}$.}
\end{figure}

\begin{figure}
\plotone{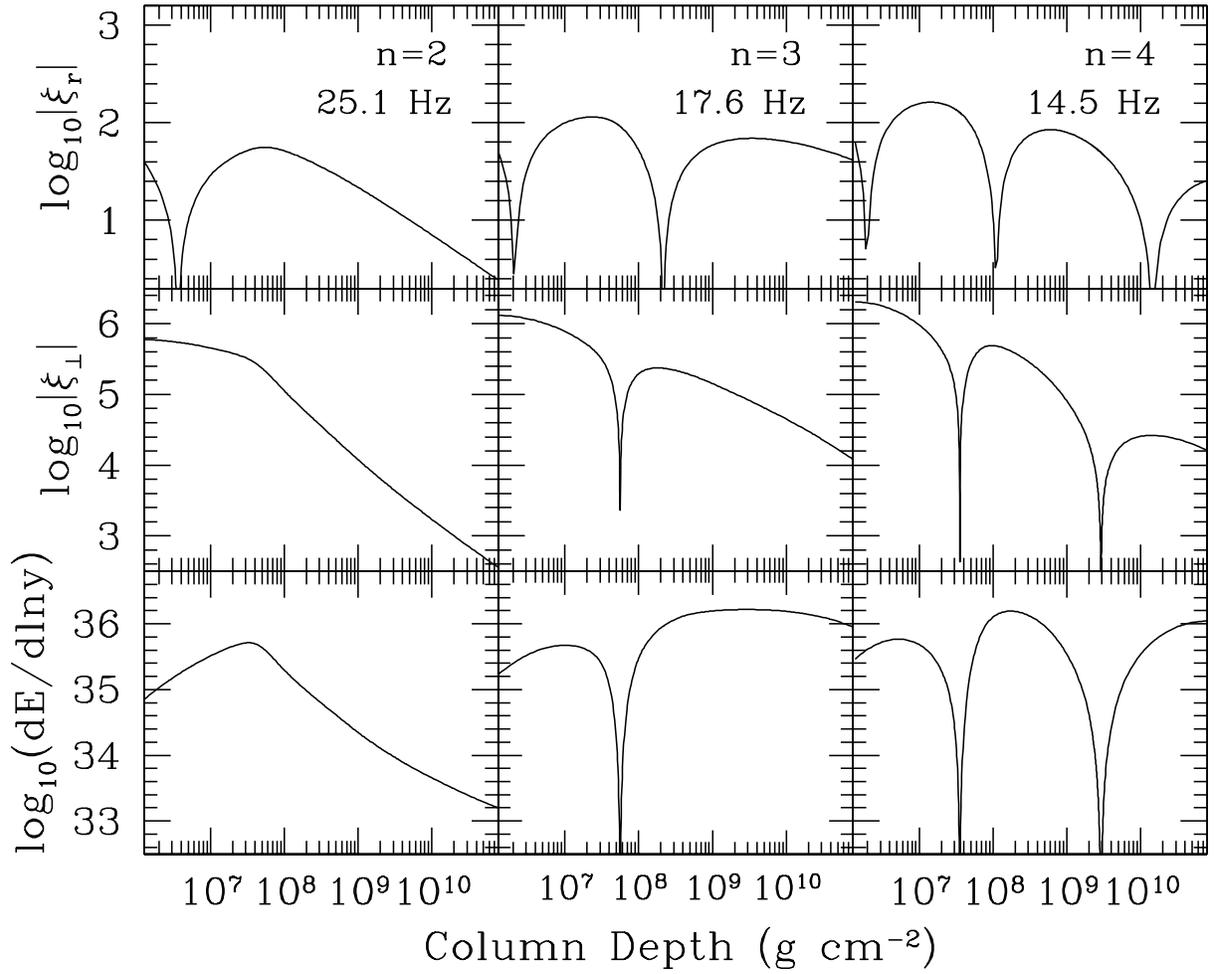}
\figcaption{The radial eigenfunctions for the first three g-modes
(with zero crossings in $\xi_r$) on
a $\dot{m}=5\dot{m}_{\rm Edd}$, $F_{\rm crust}=0$ accreting
neutron star envelope. The top (middle) panel shows the absolute
displacement of the mode amplitude $\xi_r$ ($\xi_\perp$) in units of
cm. The normalization is arbitrary, but the relative amplitudes
are significant.
The cusps are zero crossings where the amplitude changes sign.
The bottom panel shows the logarithmic energy density in units
of ergs.}
\end{figure}

\begin{figure}
\plotone{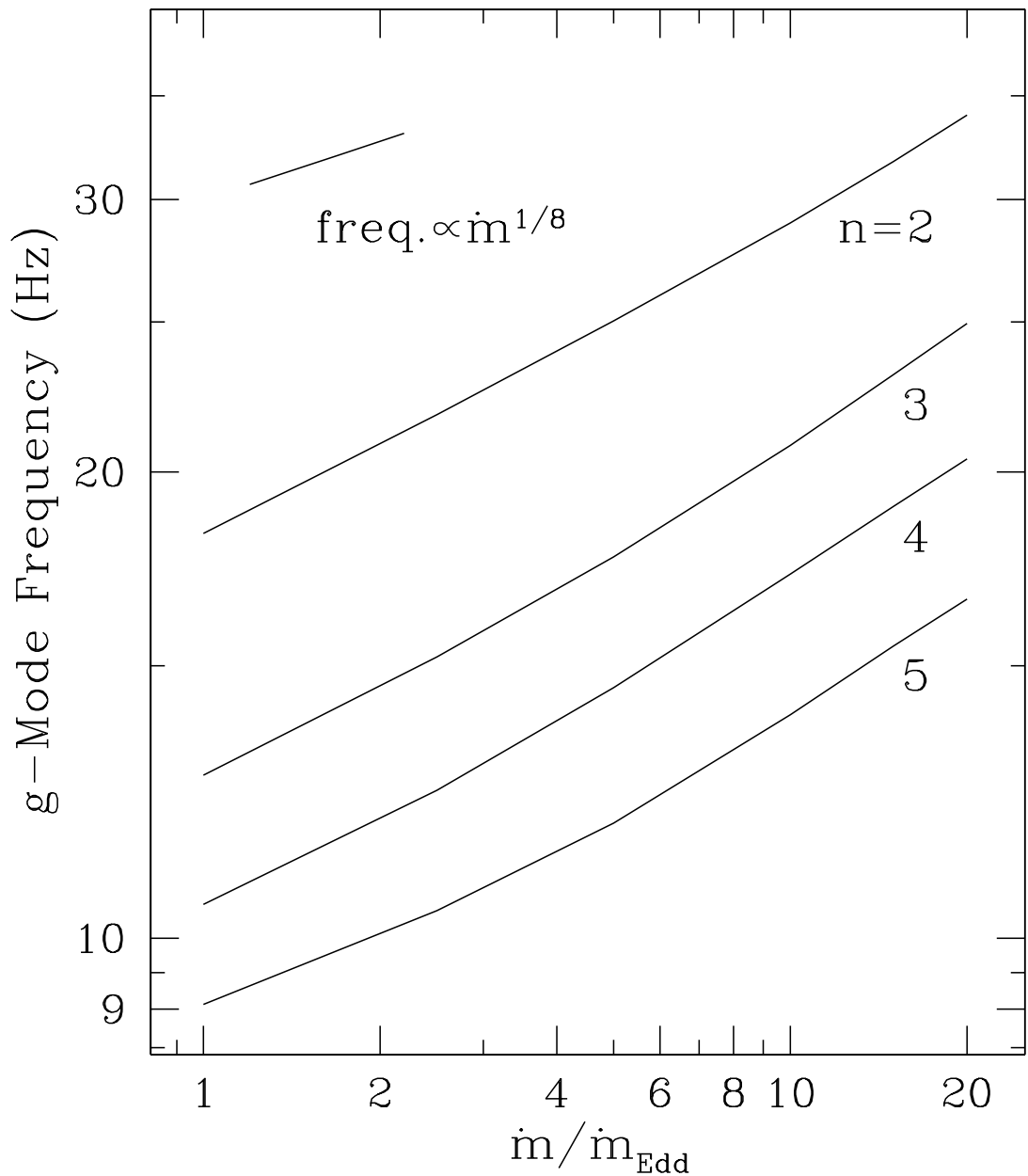}
\figcaption{The g-mode frequencies for a non-rotating neutron star
as a function of the accretion rate and for $l=1$ ($F_{\rm crust}=0$). The
frequency scales approximately $\propto \dot{m}^{1/8}$ as would be
expected from a simple analytical analysis of the mode frequency.}
\end{figure}

\begin{figure}
\plotone{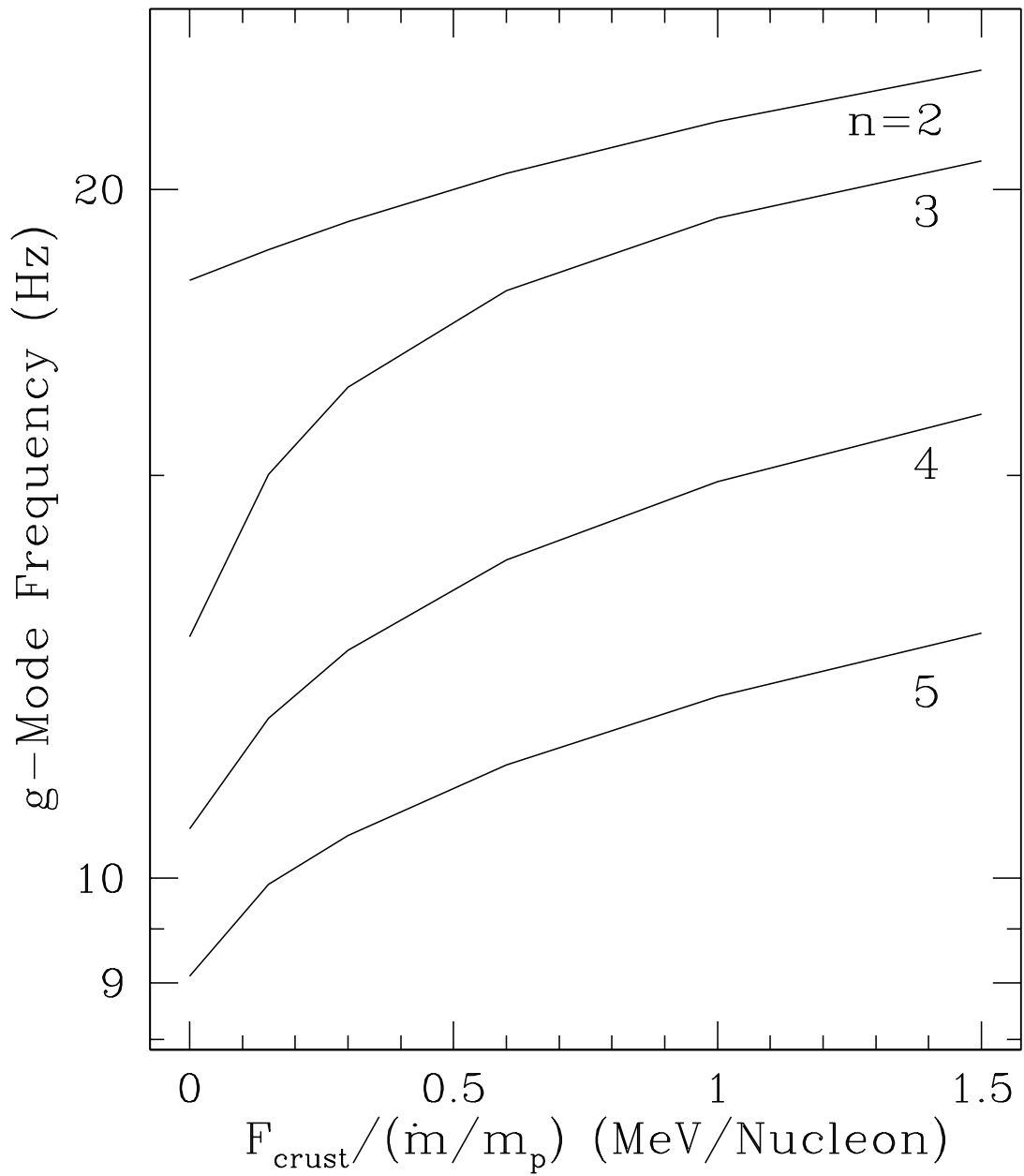}
\figcaption{Same as Figure 5, but as a function of \fc instead
($\dot{m}=\dot{m}_{\rm Edd}$). The
shallow surface wave ($n=2$) frequency does not show the same degree of
dependence on this change of the bottom boundary conditions.}
\end{figure}

\begin{figure}
\plotone{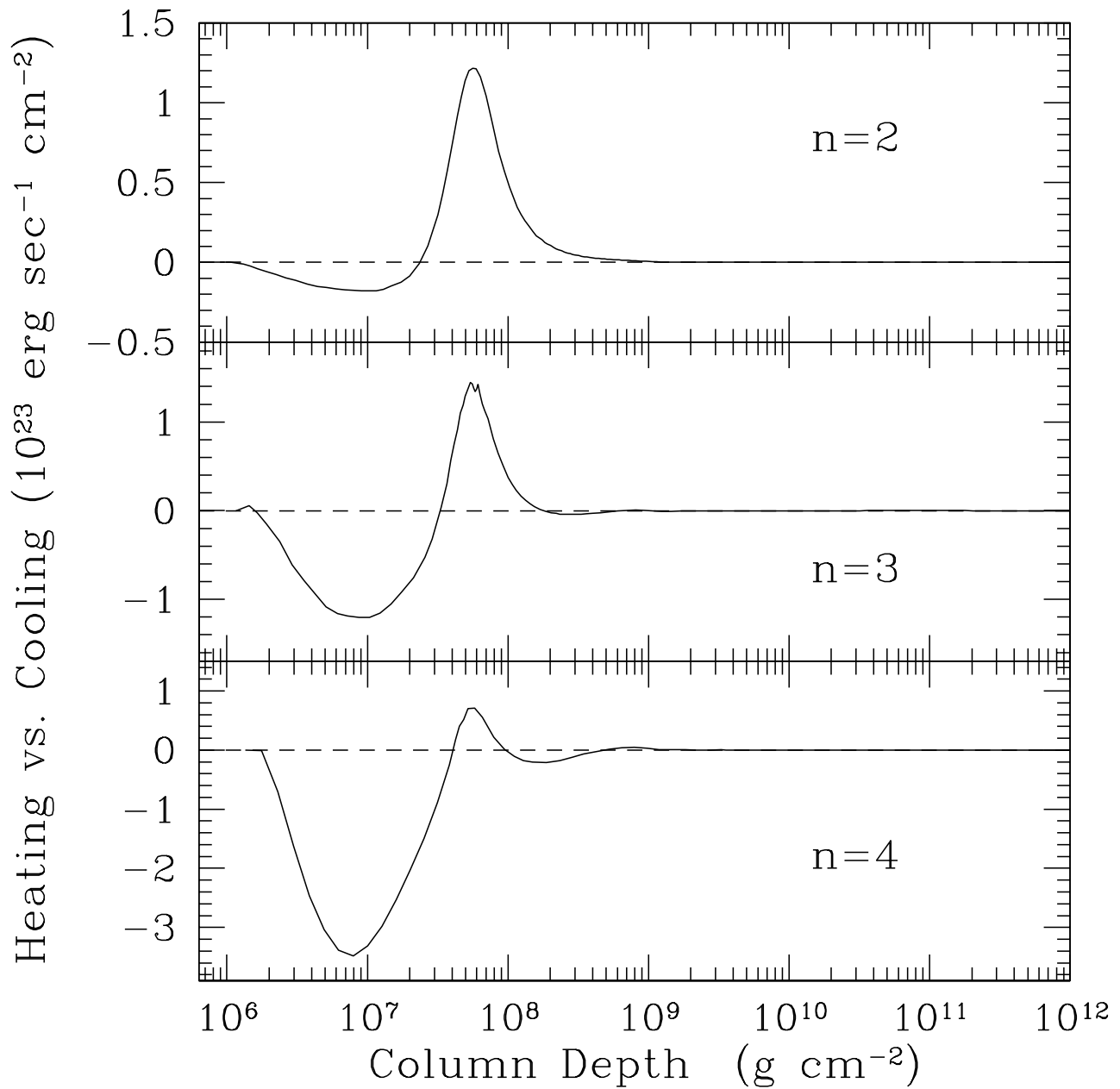}
\figcaption{The logarithmic work integral integrand, equation (26),
for three different modes. Regions
above zero are pumping the mode while regions below zero are damping the
mode. The majority of the pumping is due to increases in nuclear burning
near the burning layer ($y\approx5\times10^7\textrm{ g cm}^{-2}$), while
most of the radiative damping happens near the surface. The integrand
necessarily goes to zero at the outer boundary condition
(where $\Pi = t_{\rm th}$) because
$\Delta P/P=0$. From this figure it is clear that the shallow surface
wave ($n=2$) is excited while the others are damped.}
\end{figure}

\begin{figure}
\plotone{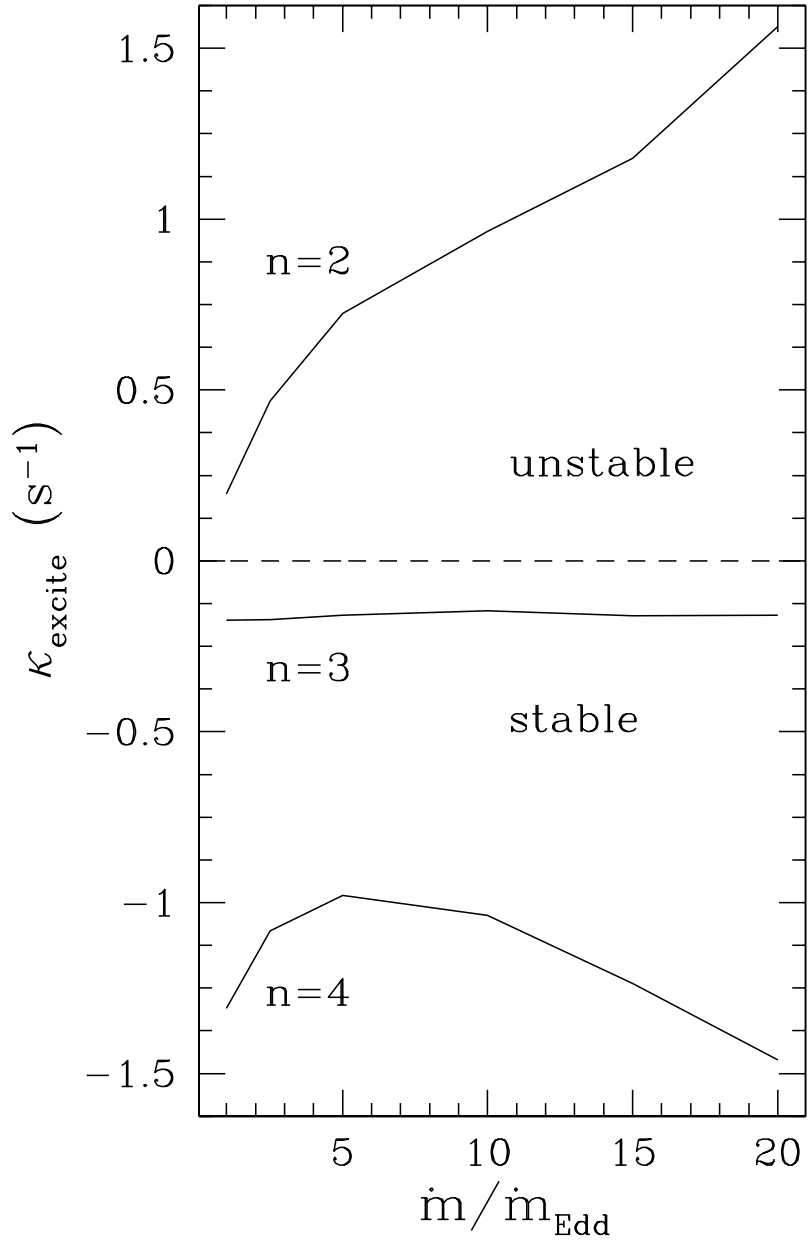}
\figcaption{The excitation or damping rate
due to nonadiabatic effects as a function of accretion rate. The inverse of
$\kappa_{\rm excite}$ is the excitation (or damping) time.
The shallow surface wave ($n=2$) is excited more rapidly as $\dot{m}$
increases while all higher order g-modes are damped.}
\end{figure}

\begin{figure}
\plotone{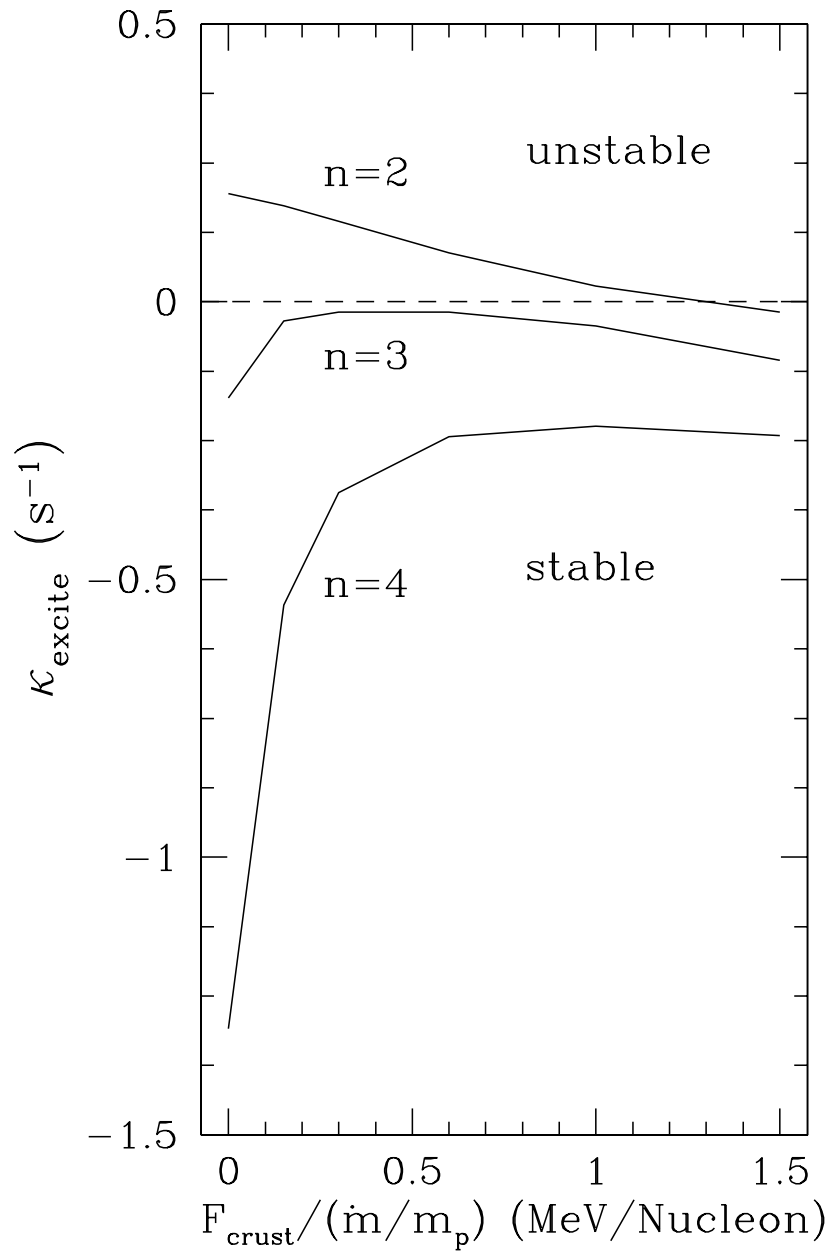}
\figcaption{Same as Figure 8, but as a function of \fc instead
($\dot{m}=\dot{m}_{\rm Edd}$). As \fc
increases, eventually even the shallow surface wave is damped
(\fc$\approx1.3 \textrm{ MeV/Nucleon}$).}
\end{figure}

\begin{figure}
\plotone{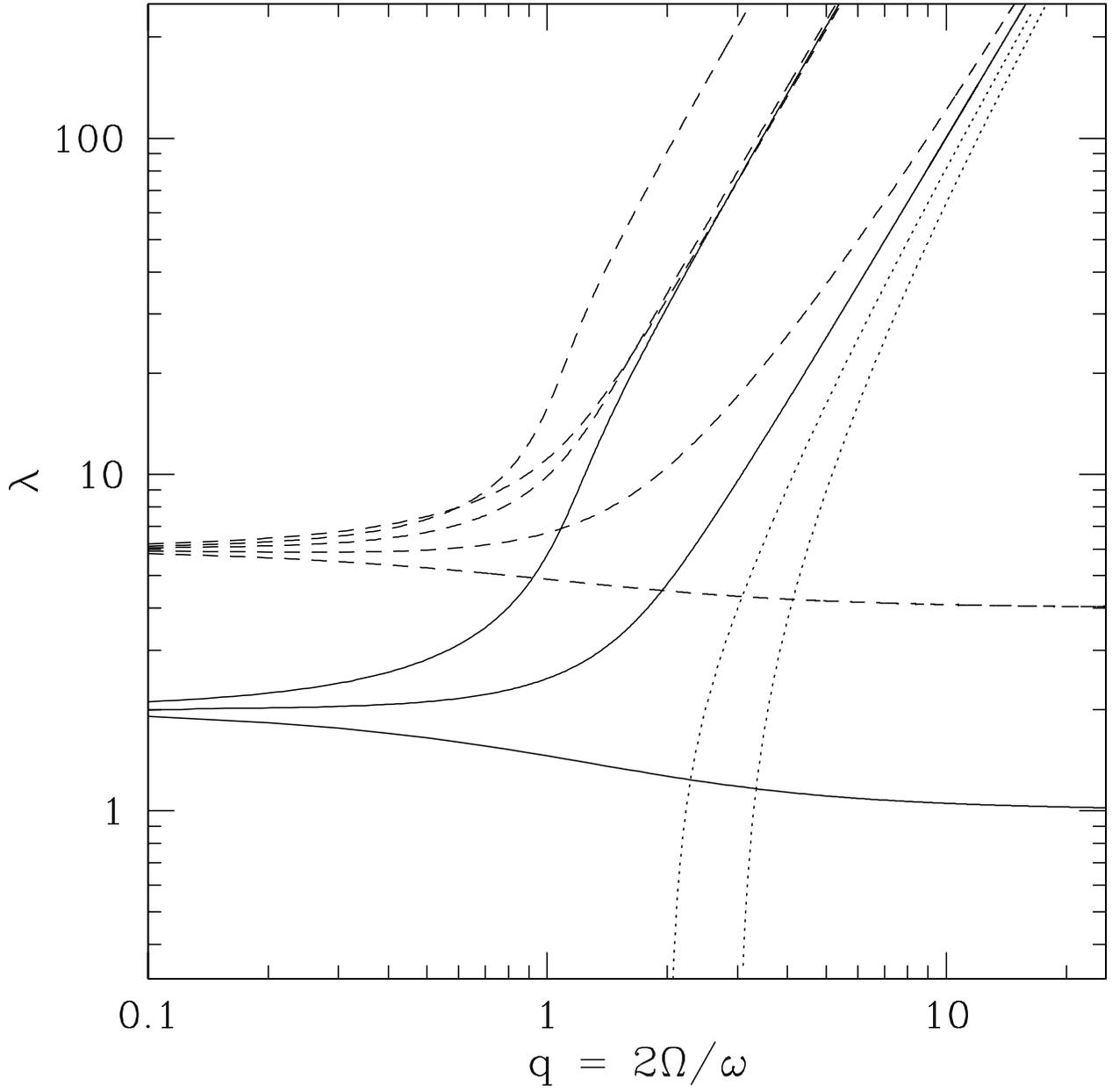}
\figcaption{The effective wavenumber, $\lambda$, for modes on a rotating
star versus the quantity $q=2\Omega/\omega$. Solid lines denote
modes for which $l=1$ in the non-rotating limit with $m$ values of -1, 0,
and 1 (from bottom to top). The dashed lines denote modes for which $l=2$
in the non-rotating limit with $m$ values -2, -1, 0, 1, and 2 (from
bottom to top). For $q\ll1$, $\lambda\approx l(l+1)$. For $q\gg1$,
$\lambda\propto q^2$, except for the lowest $m$ mode which goes to
$\lambda=m^2$ (the Kelvin mode). The dotted lines denote two of the
odd parity r-modes (from left to right, $l=1, m=1$ and $l=2, m=2$)
which satisfy $\omega = 2 m\Omega/l(l+1)$
in the $\lambda\ll1$ limit.}
\end{figure}

\begin{figure}
\plotone{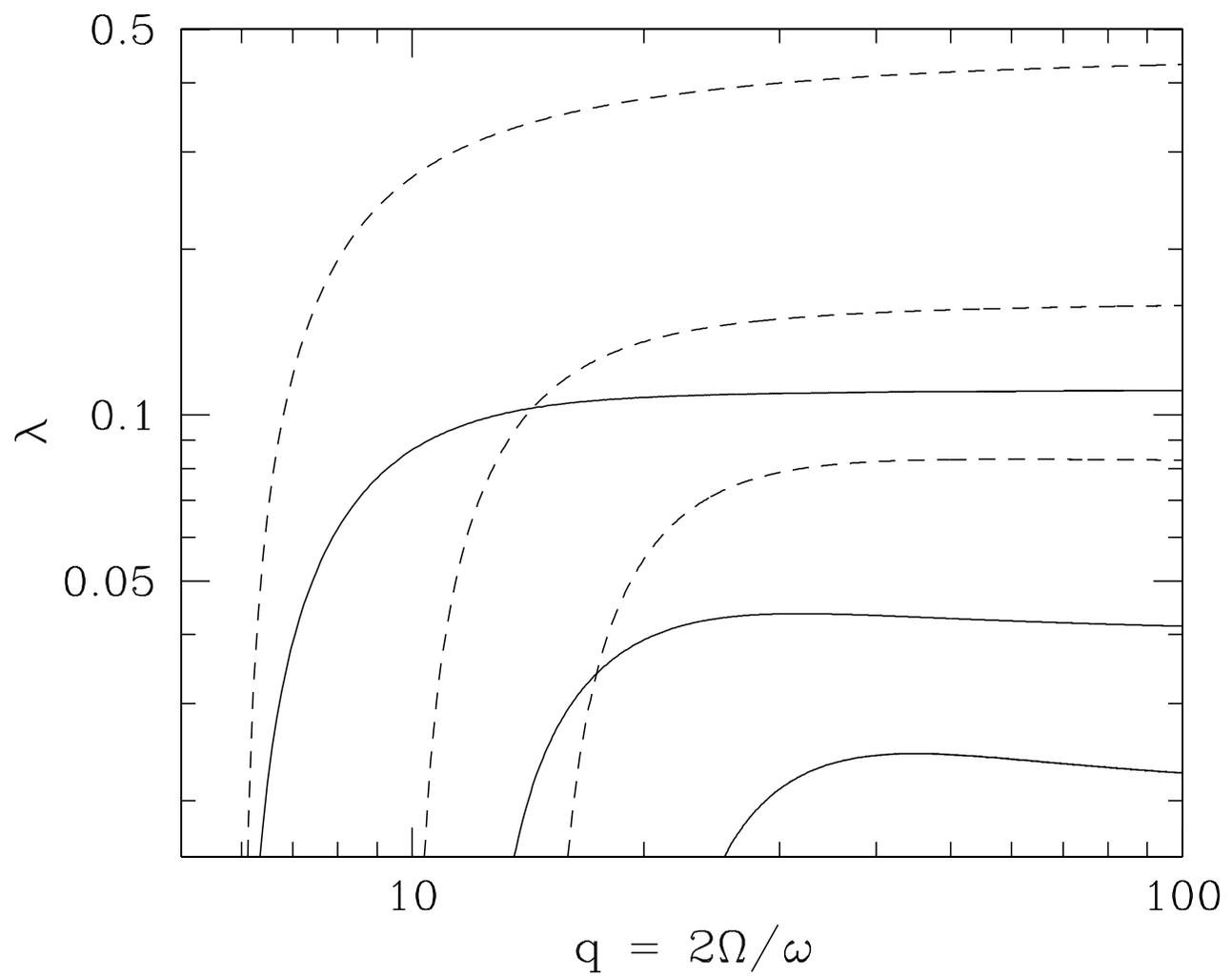}
\figcaption{The effective wavenumber, $\lambda$, for modes on a rotating
star versus the quantity $q=2\Omega/\omega$ for three of the
$m=1$ r-modes (solid lines; from left to right, $l=2$, $l=3$, and $l=4$)
and three of the $m=2$ r-modes (dashed lines; from left to
right, $l=3$, $l=4$, and $l=5$).}
\end{figure}

\begin{figure}
\plotone{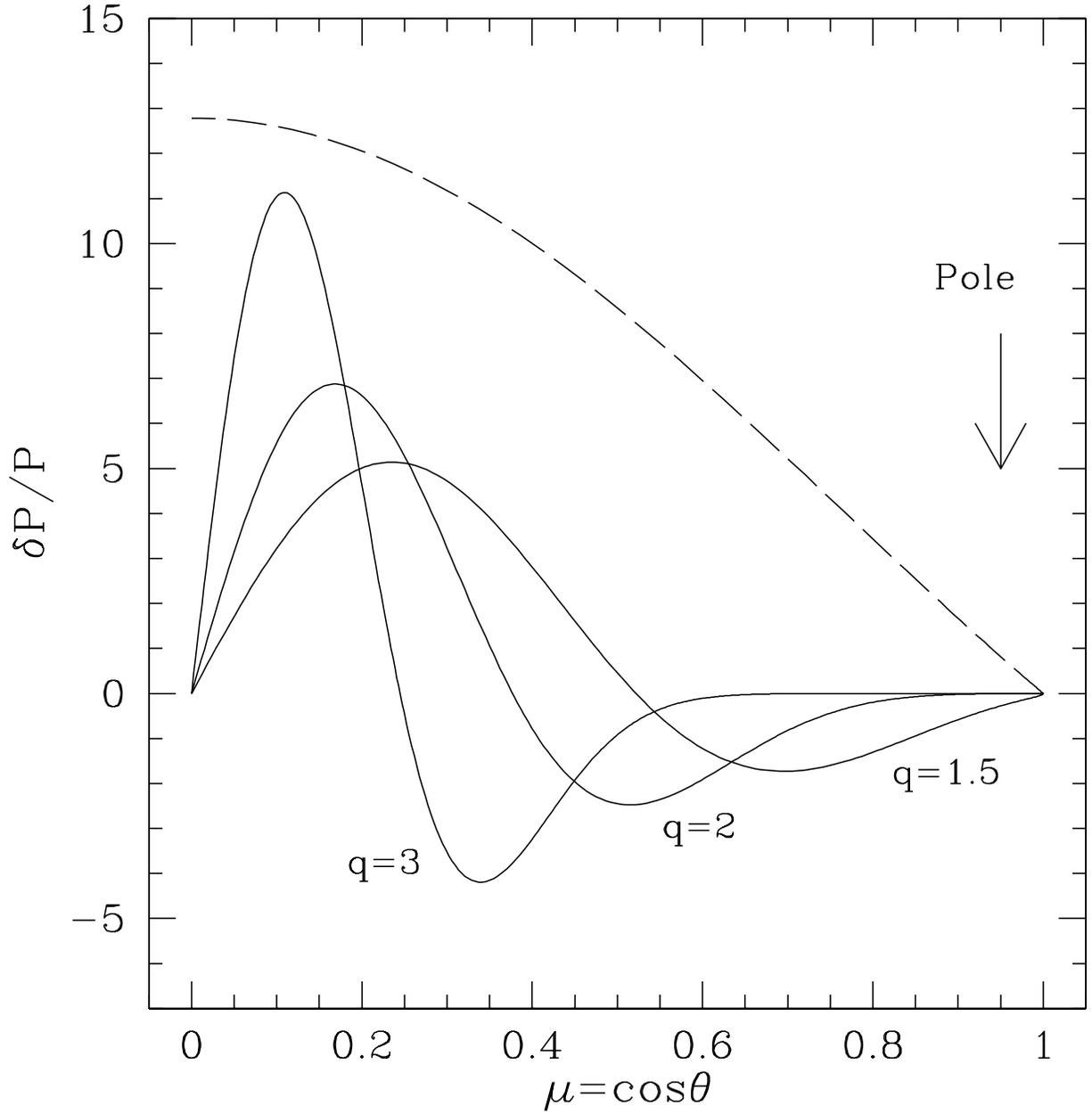}
\figcaption{Example latitudinal eigenfunctions for four different
rotationally modified modes. The solid lines denote odd parity
g-modes with $q =$ 1.5, 2.0, and 3.0.
The mode amplitude is exponentially
damped for $\mu\gtrsim1/q$ and squeezed toward the
equator as $q$ increases. The dashed line shows a characteristic
Kelvin mode which has a much greater latitudinal extent than the
others (and always has even parity). The amplitudes of these
eigenfunctions are arbitrary.}
\end{figure}

\begin{figure}
\plotone{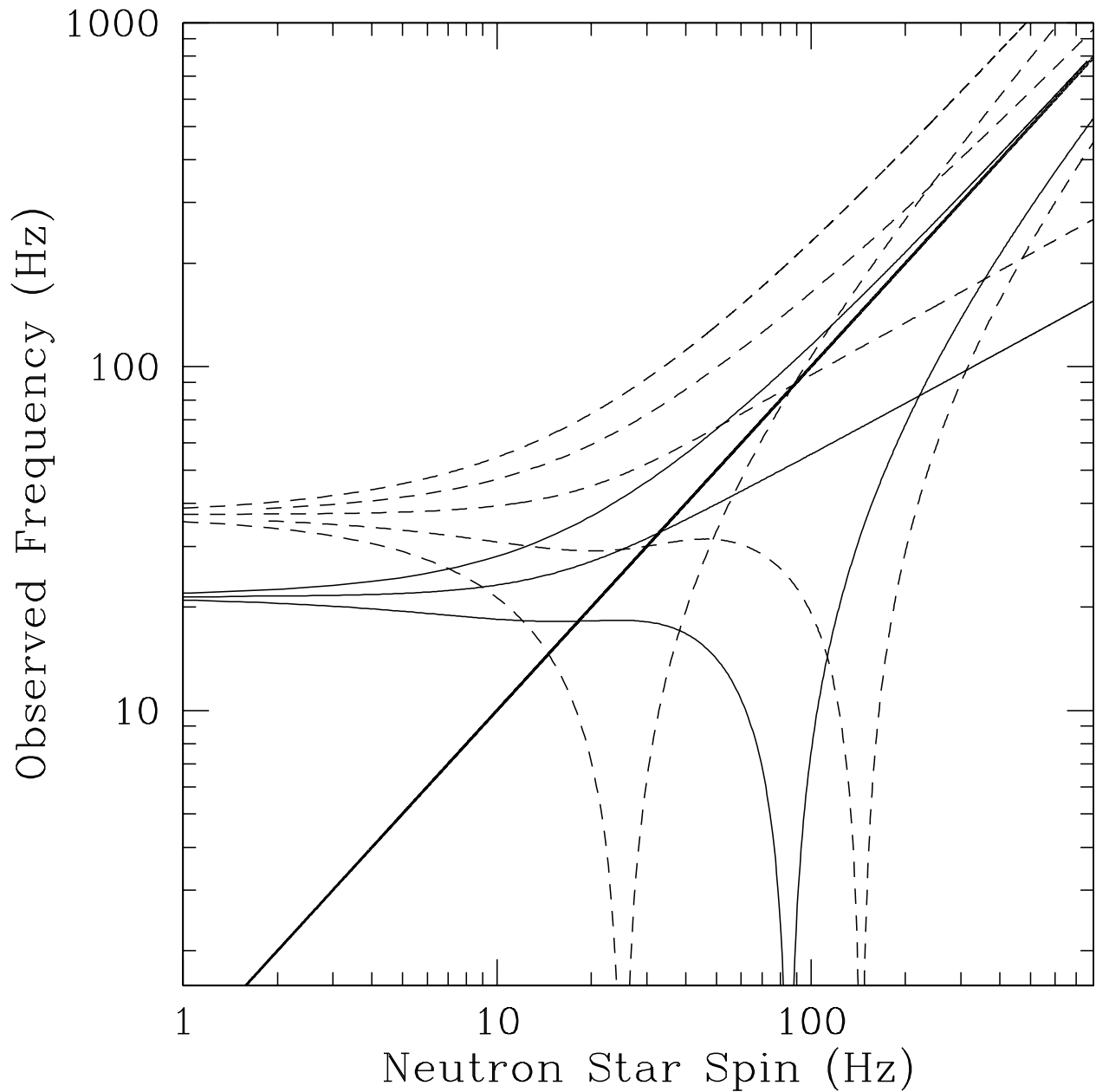}
\figcaption{A sample of observed frequencies as a function of the neutron star
spin rate for a rotationally modified $\omega_0/(2\pi)=21.4\textrm{ Hz}$
g-mode (frequency for a $\dot{m}=\dot{m}_{\rm Edd}$,
\mbox{\fc$=1.0\textrm{ MeV/Nucleon}$},
shallow surface wave in the non-rotating limit).
Solid (dashed) lines denote modes that have $l=1$ ($l=2$) in the non-rotating
limit. The thick solid line denotes the neutron star spin. The mode which
appears closest to, and above, the spin is the $m=-1$ Kelvin mode.}
\end{figure}

\begin{figure}
\plotone{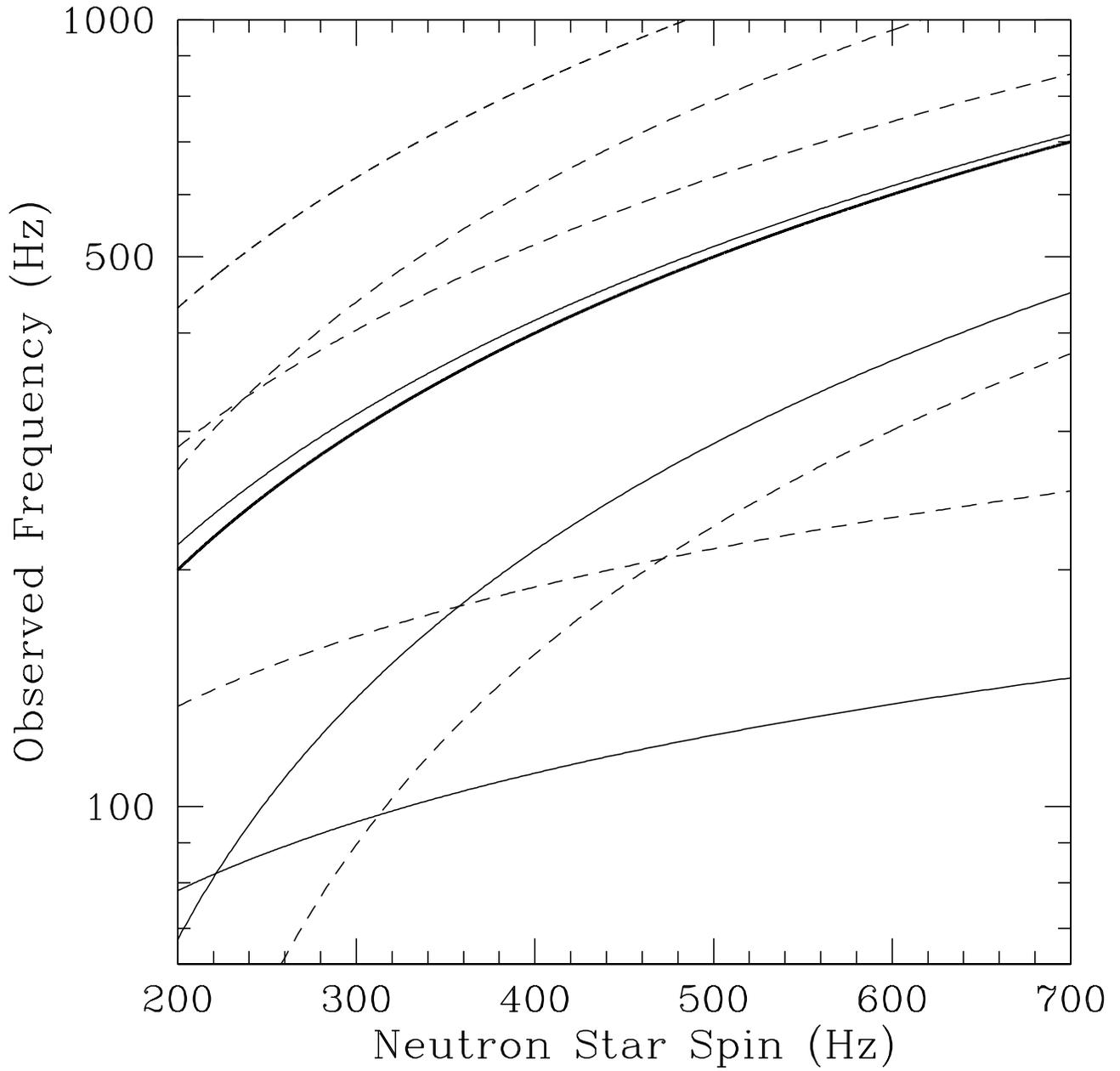}
\figcaption{Same as Figure 13, but focusing on a range of spin frequencies
that may be relevant for helium accreting neutron stars such as 4U 1820--30.
If such a pattern of frequencies were seen from a neutron star it could be
used to constrain properties of the star (e.g. spin and radius).}
\end{figure}

\end{document}